\documentclass[twocolumn,pra,aps,amsmath,amssymb,superscriptaddress,showpacs,floatfix]{revtex4-1} 
 
\usepackage[utf8]{inputenc}
\usepackage{graphicx}
\usepackage{color}

\def\be{\begin{eqnarray}}   
\def\ee{\end{eqnarray}}

\begin{document}

\title{First-principles simulations for attosecond photoelectron spectroscopy based on
time-dependent density functional theory}

\author{ Shunsuke A. Sato}
\affiliation{Max Planck Institute for the Structure and Dynamics of Matter and Center for Free Electron Laser Science, 22761 Hamburg, Germany}

\author{Hannes~H\"ubener}
\affiliation{Max Planck Institute for the Structure and Dynamics of Matter and Center for Free Electron Laser Science, 22761 Hamburg, Germany}

\author{Angel Rubio}
\affiliation{Max Planck Institute for the Structure and Dynamics of Matter and Center for Free Electron Laser Science, 22761 Hamburg, Germany}
\affiliation{Center for Computational Quantum Physics (CCQ), The Flatiron Institute, 162 Fifth avenue, New York NY 10010}
\affiliation{Nano-Bio Spectroscopy Group,  Departamento de Fisica de Materiales, Universidad del Pa\'is Vasco UPV/EHU- 20018 San Sebasti\'an, Spain}

\author{Umberto De Giovannini}
\affiliation{Max Planck Institute for the Structure and Dynamics of Matter and Center for Free Electron Laser Science, 22761 Hamburg, Germany}

\begin{abstract}
We develop a first-principles simulation method for attosecond time-resolved photoelectron spectroscopy.
This method enables us to directly simulate the whole experimental processes, including
excitation, emission and detection on equal footing.
To examine the performance of the method,  we use it to compute the reconstruction of attosecond
beating by interference of two-photon transitions (RABBITT) experiments of gas-phase Argon.
The computed RABBITT photoionization delay is in very good agreement with
recent experimental results from [Kl\"under \textit{et al}, Phys. Rev. Lett. {\bf 106} 143002 (2011)]
and [Gu\'enot \textit{et al}, Phys. Rev. A {\bf 85} 053424 (2012)].
This indicates the significance of a fully-consistent
theoretical treatment of the whole measurement process to properly describe experimental
observables in attosecond photoelectron spectroscopy.
The present framework opens the path to unravel the microscopic processes underlying
RABBITT spectra in more complex materials and nanostructures.
\end{abstract}

\maketitle

%============================================================================================================
\section{Introduction \label{sec:intro}}
%============================================================================================================

Recent progress of laser technologies has enabled the observation of ultrafast phenomena with attosecond resolution and offered novel opportunities to directly explore real-time electron dynamics in
matter \cite{hentschel2001attosecond,RevModPhys.81.163,krausz2014attosecond}.
Broadly speaking one can assign the available attosecond time-resolved measurement techniques to two major groups: one is based on all-optical measurements such as the attosecond transient absorption spectroscopy \cite{goulielmakis2010real,PhysRevLett.105.143002,PhysRevLett.106.123601}.
The other is based on photoelectron spectroscopy such as the reconstruction of attosecond beating by interference of two-photon transitions (RABBITT) \cite{Paul1689,muller2002reconstruction} and 
the attosecond streaking camera \cite{PhysRevLett.88.173903,kienberger2004atomic}.

In the past decade, attosecond transient absorption spectroscopy has been applied to atomic and molecular systems, and ultrafast electron dynamics in these relatively small systems has been intensively 
investigated both experimentally \cite{goulielmakis2010real,beck2014attosecond,warrick2016probing,reduzzi2016observation}
and theoretically \cite{PhysRevA.83.033405,CPHC:CPHC201201007}.
Recently, this technique has been extended to solid-state materials where
nonequilibrium electron dynamics has been investigated towards future applications 
such as petahertz devices \cite{Schultze1348,Lucchini916,zurch2017direct}.
Although attosecond spectroscopy of solid-state materials provides in principle a wealth of information on novel 
aspects of ultrafast dynamics, experimental results are often hard to interpret directly,
because of the strong nonlinearlity of light-matter interactions combined with the complex electronic 
structure of solids.
To extract microscopic insight from attosecond transient absorption spectra of solids,
first-principles simulations based on the density functional theory (DFT) \cite{PhysRev.136.B864,PhysRev.140.A1133} and 
the time-dependent density functional theory (TDDFT) 
\cite{PhysRevLett.52.997} have played a significant role
\cite{Schultze1348,Lucchini916,Schlaepfer2018natphys}.

Likewise, attosecond photoelectron spectroscopy has been applied to atomic systems \cite{Schultze1658,PhysRevLett.106.143002,PhysRevA.85.053424,Isinger893} as well as recently to 
solid-state materials \cite{cavalieri2007attosecond,Locher:15,neppl2015direct}. 
However, in spite of the intensive development of several \textit{ab-initio} approaches for atomic and molecular systems \cite{PhysRevA.84.061404,PhysRevA.86.061402,PhysRevA.89.033417},
similar approaches for solids and surfaces have not yet been established.
Therefore, in order to understand experiments on
such complex systems, further development of first-principles approaches is required.
In this regard one promising candidate is represented by real-time electron dynamics simulations based on TDDFT. 
Real-time TDDFT simulations of solids have been already applied to ultrafast as well as strong-field-induced phenomena such as attosecond transient absorption spectroscopy \cite{Lucchini916}, high harmonic generation \cite{doi:10.1063/1.4716192,PhysRevLett.118.087403,PhysRevA.97.011401}, laser-induced damage \cite{PhysRevB.92.205413}, and laser-induced magnetism \cite{noejung2018}.

In a recent work \cite{deGiovannini:2016bb} the authors have introduced a method to compute the angle-resolved photoelectron spectrum of solid-state systems.
This method is based on the calculation of the photoelectron flux trough a closed surface that can be simulated with TDDFT in a real-space real-time implementation.
Here we illustrate how this approach can be employed to calculate attosecond photoelectron spectra of finite systems. 
This constitutes a fundamental benchmark towards the application to solid-state materials.

For this purpose, we perform the attosecond photoelectron spectroscopy simulation of an Argon atom and compare the theoretical results with recent experiments. 
In particular here we focus on RABBITT experiments, since the alternative approach, attosecond streaking, provides equivalent information~\cite{RevModPhys.87.765,Cattaneo:16} and we emphasize that the TDDFT attosecond photoelectron spectroscopy can be straightforwardly applied also to the attosecond streaking technique.

RABBITT has been originally introduced for the temporal characterization of attosecond pulses
\cite{Paul1689}, and has then been employed to investigate the photoionization delay in atoms and molecules~\cite{PhysRevLett.106.143002,RevModPhys.87.765,Isinger893}
as well as, in recent times, in solid-state surfaces~\cite{Locher:15,Kasmi:17}.
The RABBITT technique employs two laser pulses in a pump-probe fashion: an attosecond extreme ultraviolet (EUV) pulse train is used as a pump to ionize the system while 
a femtosecond infrared (IR) pulse is used as a probe.
This configuration is designed for experiments where the pulse train is obtained by an high harmonic generation stage seeded by the IR pulse. 
As the attosecond pulse train consists of a frequency comb of odd IR frequency multiples it produces an energy comb of photoelectron spectra that are shifted by IR photons.
Probing the system with the delayed IR brings two adjacent photoelectron peaks in contact and forms an interference pattern that oscillates as a function of the delay.
This pattern encodes information on the emission delay with attosecond resolution.
In this paper we will demonstrate how the entire process can be efficiently simulated with TDDFT.

The structure of this paper is as follows: In Sec. \ref{sec:method},
we describe the theoretical and numerical methods to compute electron dynamics and photoelectron spectra
based on the TDDFT. In Sec. \ref{sec:result}, we demonstrate the first-principles simulation for attosecond photoelectron spectroscopy
and compare the theoretical results with recent experiments. We further discuss the role of a many-body effects in the photoemission process. Finally, we summarize our findings 
and provide some perspective for future work in Sec. \ref{sec:summary}.

Hartree atomic units ($\hbar=e=m_e=4\pi \epsilon_0=1$) are employed throughout the paper unless otherwise specified.

%============================================================================================================
\section{Method \label{sec:method}}
%============================================================================================================

% \subsection{Real-time real-space electron dynamics simulations based on the time-dependent density functional theory \label{subsec:tddft}}
% TDKS...

The fundamental concept of TDDFT is that all physical properties of a time-dependent system can be determined 
through their functional dependence on the time-dependent interacting many-body density \cite{PhysRevLett.52.997}, 
$n({\bf r},t)$ and the initial many body state, which can be disregarded if we start from
the ground state.
The idea of both DFT and TDDFT, is to obtain this many-body density by mapping it to the density of 
a fictitious auxiliary system of non-interacting electrons: the \emph{Kohn-Sham} (KS) system.
The dynamics of the KS system can be obtained by propagating the one-particle
equations for the orbitals $\varphi_i({\bf r},t)$ of a single Slater determinant, according to the
time-dependent KS (TDKS) equations
\begin{equation}
\label{eq:tdks}
    i\frac{\partial}{\partial t} \varphi_i({\bf r},t) = H_{\rm KS}({\bf r},t)\varphi_i({\bf r},t) ,\;\;\; i=1,\dots,N/2\,.
\end{equation}
To simplify notation we here only consider systems with an even number
of electrons $N$, so that each spatial orbital $\varphi_i$ is doubly occupied with two electrons of opposite spin.
The KS Hamiltonian governing the dynamics of the orbitals in~(\ref{eq:tdks}) is defined as:
\begin{eqnarray}
    H_{\rm KS}({\bf r},t)&=& \frac{1}{2} \left(-i \nabla + \frac{\mathbf{A}(t)}{c}\right)^2 + v_{\rm KS}[n]({\bf r},t)\,, \label{eq:hks} \\
    v_{\rm KS}[n]({\bf r},t) &=& v_{\rm ion}({\bf r},t) +v_{\rm H}[n]({\bf r},t) + v_{\rm xc}[n]({\bf r},t)\,\label{eq:vks}
\end{eqnarray}
where, due to the action of the KS potential $v_{\rm KS}$, the time-dependent density $n({\bf r},t) = 2\sum_{i=1}^{N/2}\vert\varphi_i({\bf r},t)\vert^2$ corresponds both to the real and to the KS system.
The KS potential is composed of three terms. 
The first term is the electron-ion potential provided by the nuclei, while
the second term is the electrostatic potential generated by the electronic charge density $v_{\rm H}[n]({\bf r},t) = \int\!\!{\rm d}{\bf r}'\;n({\bf r}',t)/\vert {\bf r}-{\bf r}'\vert$.
The last term $v_{\rm xc}[n]({\bf r},t)$ is the so-called exchange and correlation (xc) potential that accounts for the many-body effects deriving from the electron-electron interaction; it is a functional of the density at all times $n({\bf r},t)$ and, since its explicit form is unknown, it must be approximated. 
In this work we employ the adiabatic local-density approximation (ALDA) \cite{PhysRevLett.45.566,PhysRevB.23.5048} which is based on
the xc potential of a homogeneous electron gas evaluated with the instantaneous density in time at every
point in space. 
In order to compensate the self-interaction error \cite{PhysRevB.23.5048} of the local approximation and obtain a correct ionization 
potential we employ the simplest scheme based on the averaged-density self-interaction correction (ADSIC)~\cite{Legrand:2002jf}.

Given the energy range of the lasers employed in the simulations, it is well justified to invoke the dipole approximation for the light-matter interaction. 
Under this condition, the coupling with the laser field can be expressed in the velocity gauge which amounts to modifying the kinetic operator by adding the spatially homogeneous time-dependent vector potential $\mathbf{A}(t)$ of the classical laser field, as in Eq.~(\ref{eq:hks})\footnotemark.
The time profile of $\mathbf{A}(t)$ can accommodate any linear combination of laser fields and is therefore 
naturally suited to describe any kind of pump-probe configuration, including the one employed for RABBITT 
experiments.

\footnotetext{Note that Eq. (\ref{eq:hks}) and Eq. (\ref{eq:vks}) correctly describe the coupling with external electric fields, but neglect contributions from the magnetic fields. To correctly account for magnetic fields one would need to resort to a current-density functional theory formulation of the problem where exchange and correlation are expressed via a vector potential ${\bf A}_{xc}[{\bf j}]$. 
However, the effect of the magnetic component of a laser electromagnetic radiation is much smaller than that of the electric component and can be safely neglected in the presently discussed context.}

To obtain the photoelectron spectrum from the time-dependent KS orbitals, we use the fact that it can be expressed as a flux integral of the ionization current through a closed surface.
This approach is based on the t-SURFF method, first introduced by Scrinzi \cite{Tao:2012ev} for one-electron
systems and later extended to many electrons with TDDFT \cite{Wopperer:2017bm}.
According to this formulation the momentum-resolved photoelectron probability $\mathcal{P}({\bf p})$, i.e. the probability to measure an electron with a given momentum $\mathbf{p}$, can be expressed as  
\begin{equation}\label{eq:tsurfPp}
\mathcal{P}({\bf p}) =  
 \frac{2}{N}\sum_{i=1}^{N/2}  \left|  \int_0^\infty {\rm d }\tau \oint_S {\rm d } {\bf s}\cdot \langle \chi_{\bf p}(\tau) | \hat{\bf j} |\varphi_i(\tau) \rangle \right|^2
\end{equation}
where $\hat{\bf j}$ is the single-particle gauge-invariant current density operator and $\chi_{\bf p}({\bf r},t)= (2\pi)^{-\frac{3}{2}} e^{i ({\bf p}+{\bf A}(t))\cdot{\bf r}} e^{i \Phi({\bf p},t)}$ . The phases $\Phi({\bf p},t)=\int_0^t {\rm d}\tau \frac{1}{2} \left( {\bf p} + \frac{{\bf A}(t)}{c} \right)^2$ 
describe Volkov waves of 
momentum $\mathbf{p}$ that are the analytical solutions of the time-dependent Schr\"odinger equation 
for free particles in a field.  
The bracket notation in the equation is thus used as shorthand to indicate the 
evaluation of the current-density operator matrix element between KS orbitals and Volkov waves. 
The energy-resolved photoelectron spectrum,  $\mathcal{P}(E)$, employed to build the RABBITT traces can be obtained by integrating the angular dependence of $\mathcal{P}({\bf p})$ as follows
\begin{equation}
   \mathcal{P}(E)=\int_0^{4\pi}{\rm d}\Omega\,  \mathcal{P}(E=\frac{{\bf p}^2}{2},\Omega) \,.
   \label{eq:integrated-pes}
\end{equation}

\begin{figure}[htbp]
%\centering
\includegraphics[width=0.84\columnwidth]{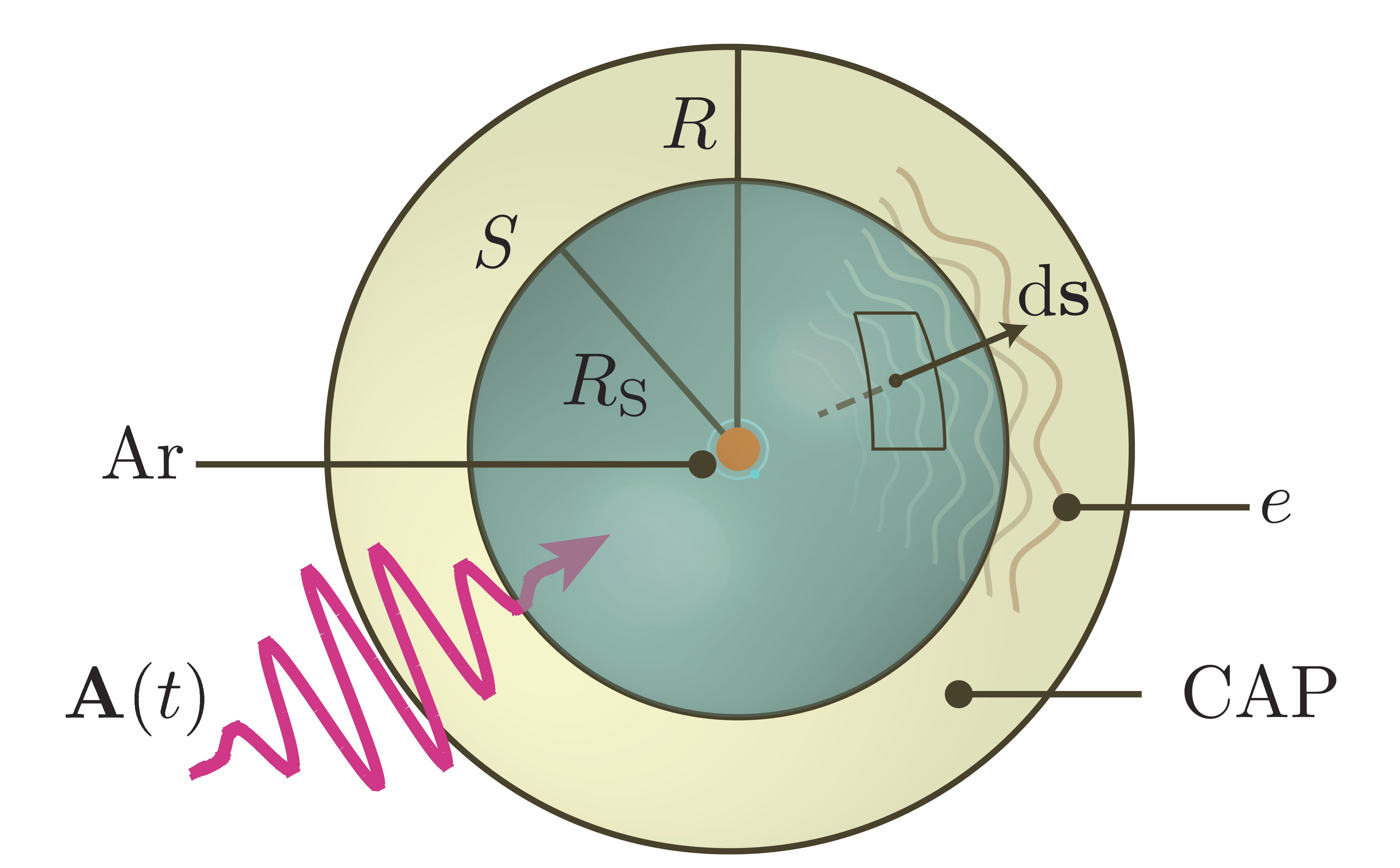}
\caption{\label{fig:tsurff} Scheme illustrating the geometry employed to calculated the photoelectron
spectrum with t-SURFF and TDDFT.}
\end{figure}

This approach to photoemission is particularly suited to numerical implementations where the TDKS 
equations~(\ref{eq:tdks}) are solved in real-space and propagated in real-time. 
In our implementation the spatial coordinates are discretized on a cartesian grid with spacing 
$\Delta =0.3$~a.u. and the equations are solved on a spherical box of radius $R=30$~a.u.. 
The TDKS equations are propagated under the influence of a time dependent field with a time step $\Delta 
t=0.04$~a.u. starting from the ground state configuration. 
The photoelectron probability is calculated with~(\ref{eq:tsurfPp}) by collecting the flux integral calculated on 
a spherical surface of radius $R_{\rm S}=20$~a.u. while the KS orbitals are propagated over time.
To prevent spurious reflections from the boundaries of the simulation box we employ a complex absorbing potential 
(CAP) acting on the region outside the surface $S$ with parameters tuned in such a way to be maximally efficient 
in the energy region where we expect photoelectrons to be mostly emitted~\cite{DeGiovannini:2015jt}.
The geometry employed in the simulations is summarized in Fig.~\ref{fig:tsurff}.

Finally, since the inner shell electrons of Argon are not expected to take significant part in the ionization 
dynamics, we use the Hartwigsen-Goedecker-Hutter (HGH) pseudopotential~\cite{PhysRevB.58.3641} that effectively accounts for the core electrons and consider 
explicitly only the $n=3$ electrons.

All the simulations presented are carried out with the \emph{Octopus} code~\cite{Strubbe:2015iz}.

%============================================================================================================
\section{Result \label{sec:result}}
%============================================================================================================

In this section, we examine the performance of the TDDFT simulation for the attosecond 
photoelectron spectroscopy. For this purpose, we simulate the RABBITT measurement processes for an Argon atom.
We first explain how to simulate the entire the RABBITT measurement. Then, we compare the theoretical results with the recent experimental data \cite{PhysRevLett.106.143002,PhysRevA.85.053424}.
Finally, we investigate the role of many-body effects in the RABBITT photoemission delay.

\subsection{RABBITT spectroscopy \label{subsec:RABBITT-tddft}}

Here, we revisit the RABBITT pump-probe technique from a computational point of view.
The RABBITT measurement is a pump-probe experiment that employs 
an attosecond EUV pulse train as a pump and an IR femtosecond pulse as a probe. Importantly, the attosecond pulse train 
is generated by high-order harmonic generation of the same IR femtosecond pulse.
Therefore, the attosecond EUV pulse train consists of odd harmonics of the IR field.

In this work, we employ the following form for the femtosecond IR pulse,
\be
A_{IR}(t)= -\frac{c E_{IR}}{\omega_{0}} \left[\cos\left ( \frac{\pi t}{T_{IR}}\right)\right]^2\sin \left(\omega_{0}t\right),
\label{eq:IR-pulse}
\ee
in the domain $-T_{IR}/2<t<T_{IR}/2$ and zero outside. 
Here $\omega_0$ is a mean frequency of the IR pulse, and $T_{IR}$ is the full duration of the pulse.
The maximum field amplitude $E_{IR}$ is related to the peak laser intensity as $I_{IR}=cE^2_{IR}/8\pi$.
We set $\omega_0$ to $1.55$ eV$ $, and $T_{IR}$ to $30$ fs. The peak intensity $I_{IR}$ is set to
$10^{11}$ W/cm$^2$.
As a corresponding attosecond pulse train, we employ the following form:
\be
A_{EUV}(t)=-\frac{c E_{EUV}}{\omega_{EUV}} \cos^4 \left(\frac{\pi t}{T_{train}}\right)
\cos^6 \left( \omega_0 t\right) \sin\left(\omega_{EUV} t \right), \nonumber \\
\label{eq:EUV-pulse}
\ee
in the domain $-T_{train}/2<t<T_{train}/2$ and zero outside. Here $\omega_{EUV}$ is the central frequency, 
and $T_{train}$ is the full duration of the pulse train.
We set $\omega_{EUV}$ to $25\omega_0$, and $T_{train}$ to $10$ fs. We also set the peak laser intensity $I_{EUV}=cE^2_{EUV}/8\pi$
to $5\times 10^{10}$ W/cm$^2$.

\begin{figure}[htbp]
%\centering
\includegraphics[width=0.84\columnwidth]{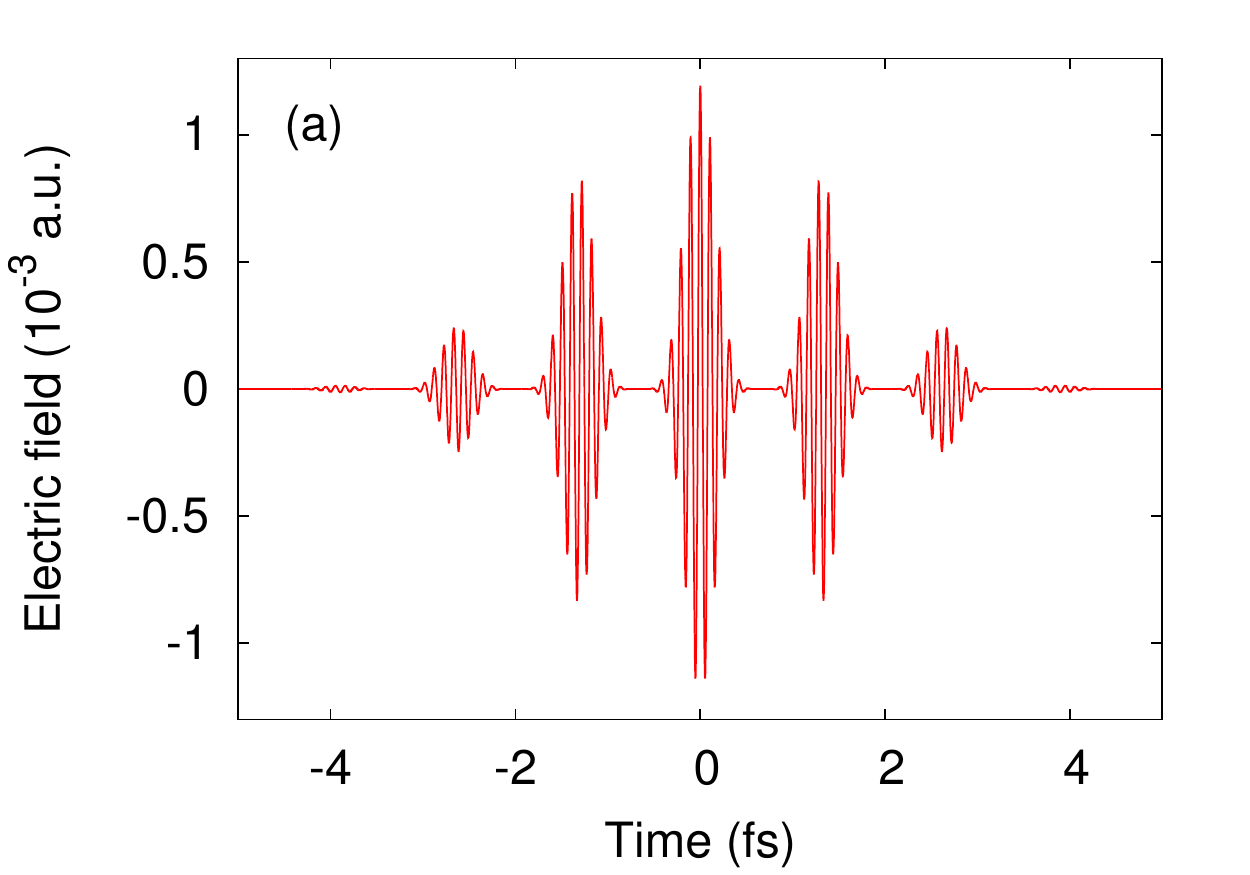}
\includegraphics[width=0.8\columnwidth]{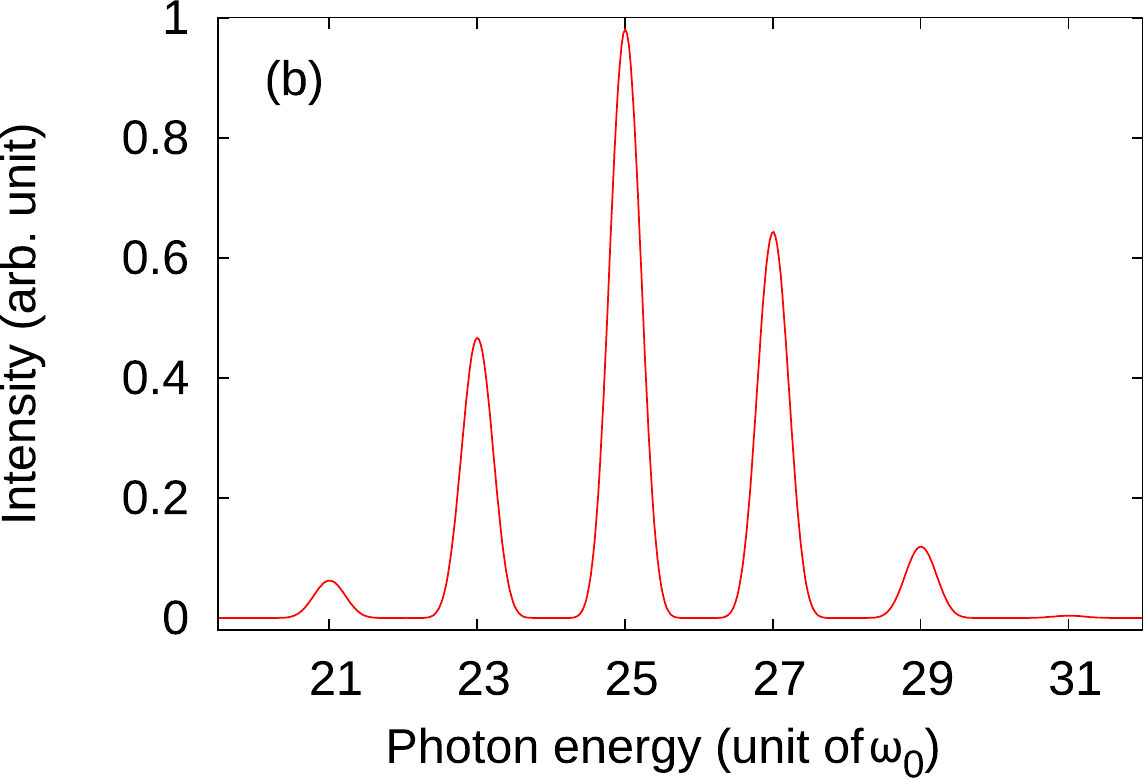}
\caption{\label{fig:Etw_ATPs} Profile of the attosecond pulse train in time-domain (a)
and the frequency domain (b).}
\end{figure}

Figure \ref{fig:Etw_ATPs} shows the attosecond pulse train of Eq. (\ref{eq:EUV-pulse}) 
in the time-domain (a) and the frequency-domain (b). As seen from Fig. \ref{fig:Etw_ATPs} (a), 
several attosecond pulses follow each other in a line with equal distance in time-domain. Each pulse has
about $120$ attoseconds full-width half-maximum.
This train forms a comb in the frequency domain, as seen in Fig. \ref{fig:Etw_ATPs} (b).
We note that the comb consists of the odd order harmonics of the IR probe pulse.

\begin{figure}[htbp]
%\centering
\includegraphics[width=0.8\columnwidth]{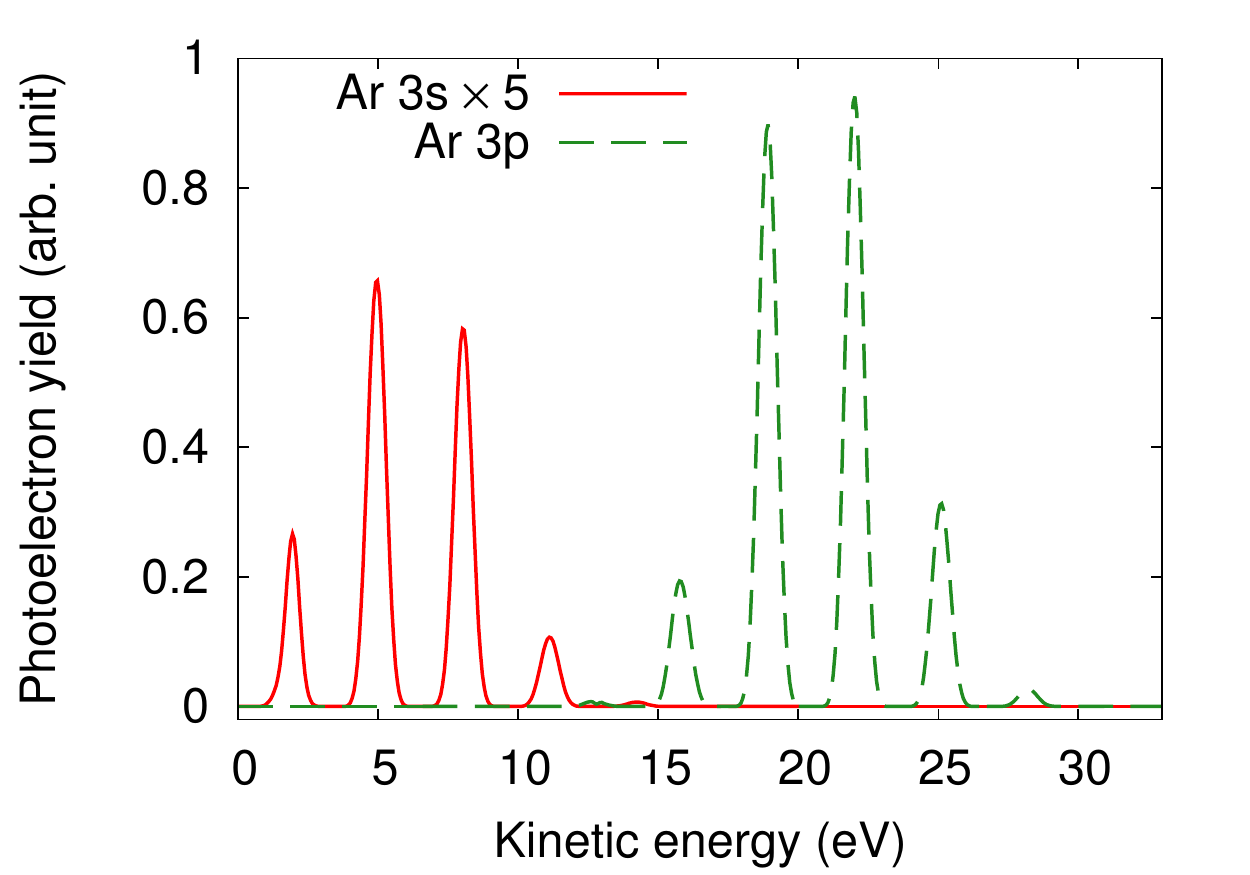}
\caption{\label{fig:PES_ATPs}
Photoelectron spectrum induced by the attosecond pulse train of Figure \ref{fig:Etw_ATPs}. 
The contribution from the Ar 3$s$ shell
is shown as a red-solid line, while that from the Ar 3$p$ shell is shown as a green-dashed line.
The contribution from the Ar 3$s$ shell is scaled by a factor of $5$.
}
\end{figure}

We then compute the photoelectron spectrum induced by the attosecond pulse train.
Figure \ref{fig:PES_ATPs} shows the photoelectron spectrum as a function of kinetic energy
of emitted electrons.
Since the photoelectron spectrum is computed based on the time-dependent Kohn-Sham orbitals,
one can naturally decompose the signal into each orbital contribution. 
In Fig. \ref{fig:PES_ATPs}, the contribution from the Ar 3$s$ shell is shown as a red-solid line, 
while that from the Ar 3$p$ shell is shown as a green-dashed line.
One sees that the two contributions are energetically well separated because of the large
difference between the ionization potentials of the 3$s$ and 3$p$ shells. 
Each contribution shows the comb structure, reflecting the frequency comb feature of
the attosecond pulse train in Fig. \ref{fig:Etw_ATPs} (b).

\begin{figure}[htbp]
%\centering
\includegraphics[width=0.8\columnwidth]{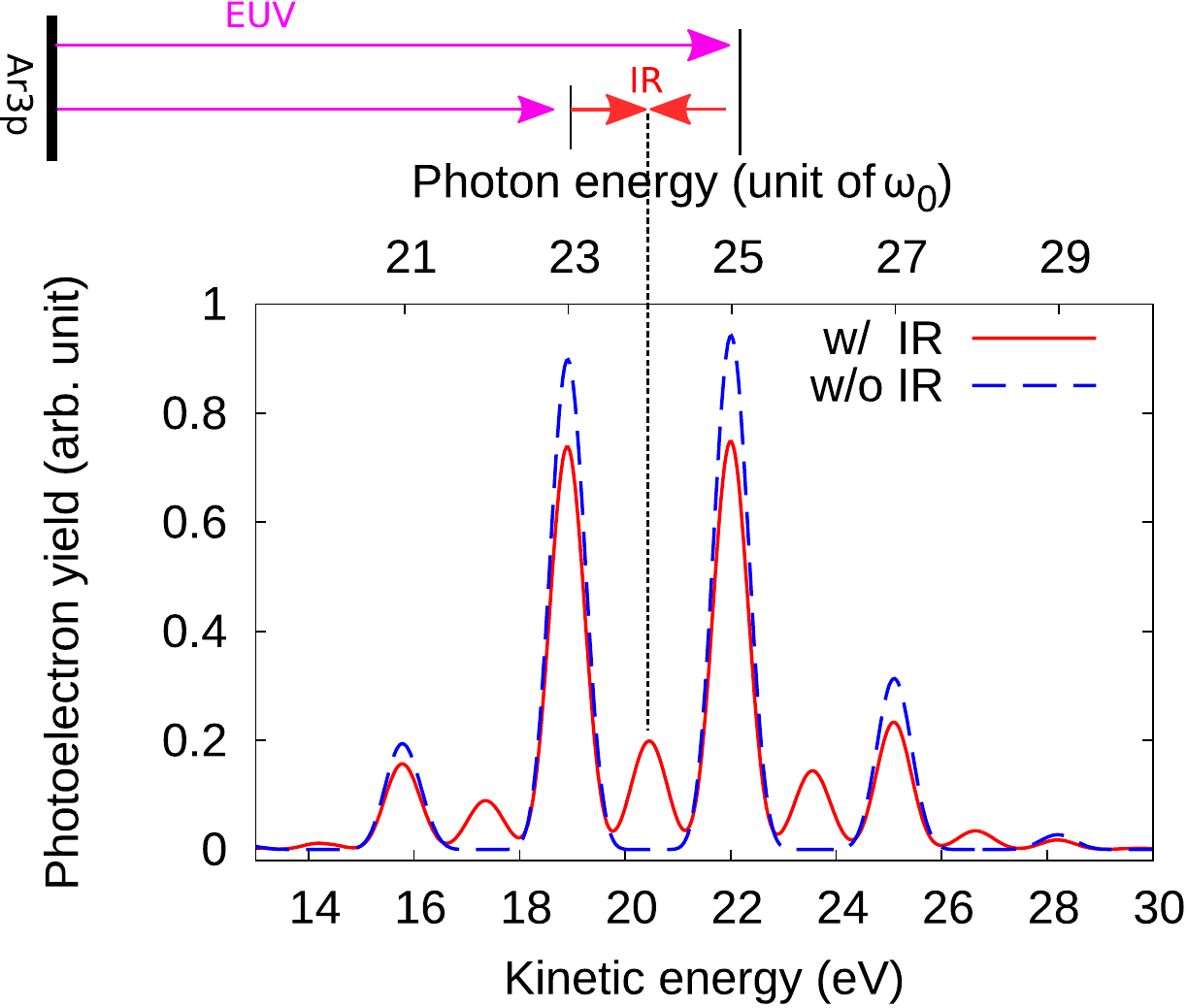}
\caption{\label{fig:PES_ATPs_IR}
Photoelectron spectra from the Ar 3$p$ shell. Red-solid line shows the result with both 
the attosecond pulse train and the IR femtosecond pulse, while the blue-dashed line
shows the result only with the pump pulse train.
}
\end{figure}

In a RABBITT experiment, the photoelectron spectrum under both the attosecond pulse train
and the femtosecond IR pulse is measured. In perfect analogy, we can compute in the theoretical simulation
the photoelectron spectrum under both the attosecond pulse train
and the IR femtosecond laser pulse. Figure \ref{fig:PES_ATPs_IR} shows 
the photoelectron spectrum from the Ar 3$p$ shell. Red-solid line shows the photoelectron spectrum
created by both the attosecond pulse train and the IR femtosecond pulse, while
blue-dashed line shows the signal solely due to the pulse train.
One sees that the IR pulse results in additional peaks between those peaks that were created only by
the pulse train. These additional peaks originate from a two-photon absorption
process: one-photon from the attosecond pulse train and the other from the IR pulse.
Adding the ionization potential of the Ar 3$p$ shell to the photoelectron kinetic energy, 
the absorbed photon energy can be calculated. The calculated absorbed photon energy
is shown as the secondary $x$-aixis of Fig. \ref{fig:PES_ATPs_IR}.
Each additional peak due to the IR field consists of two excitation paths: 
one corresponds to the EUV photon energy plus the IR photon energy, 
while the other corresponds to the EUV minus IR photon energy.
For example, as seen in the schematic picture of Fig. \ref{fig:PES_ATPs_IR}, the additional peak at the energy of $24  \omega_0$ is created by the following two
excitation paths: One is the 23rd harmonics plus the IR photon energy, while the other is
the 25th harmonics minus the IR photon energy. As discussed above, this interference between the two excitation path is the central effect used in which RABBIT spectroscopy.

\begin{figure}[htbp]
%\centering
\includegraphics[width=0.8\columnwidth]{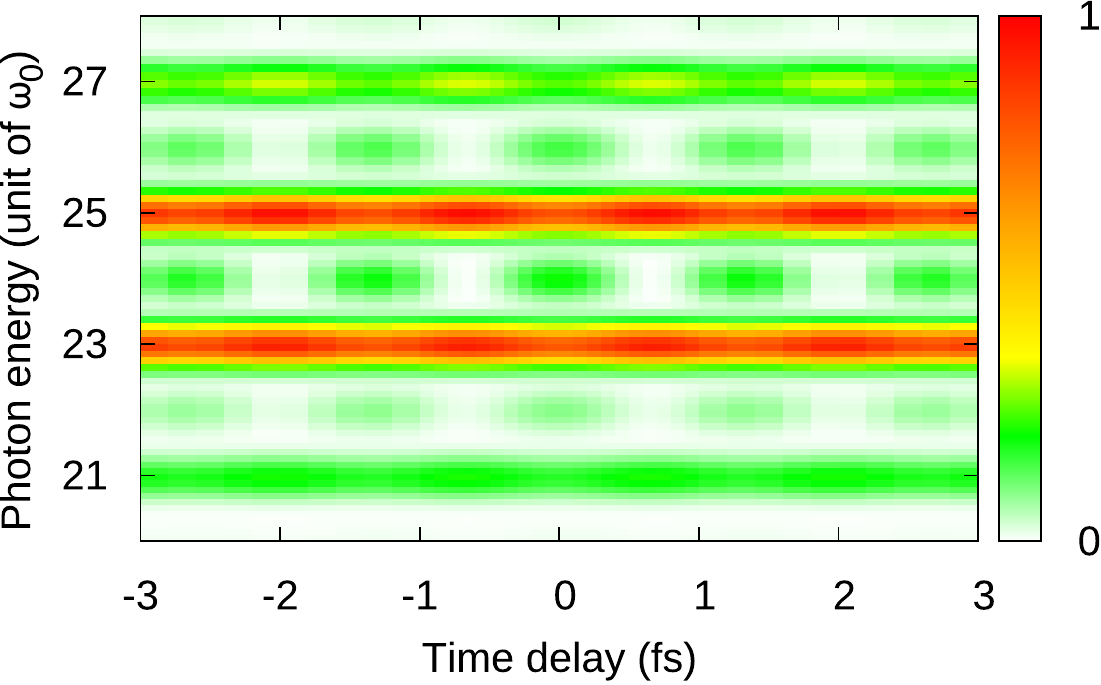}
\caption{\label{fig:RABBITT_Ar_3p}
Calculated RABBITT trace from Eq. (\ref{eq:integrated-pes}) for the Ar 3$p$ shell
using the laser pulses of Fig. \ref{fig:Etw_ATPs}.}
\end{figure}

We next perform the RABBITT pump-probe simulations by changing the time delay
between the attosecond pulse train and the IR pulse.
Figure \ref{fig:RABBITT_Ar_3p} shows the calculated photoelectron spectrum 
as a function of the time delay.
One sees that the even order side bands show an oscillating feature in time delay,
reflecting the interference of the two different two-photon absorption paths,
which is described in the schematic picture of Fig. \ref{fig:PES_ATPs_IR}.
The frequency of the oscillation is twice that of the IR frequency $\omega_0$.
Generally, each side band has its own time-delay with respect to the IR field.
Because the difference of the delay of these side bands reflects the difference of 
the photoionization delay in each excitation channel, 
the time delay in the RABBITT trace has been used to investigate the photoionization processes  \cite{PhysRevLett.106.143002,PhysRevA.85.053424,Isinger893}.
Since TDDFT can directly simulate the whole RABBITT experimental process and
provide the resulting RABBITT trace as seen in Fig. \ref{fig:RABBITT_Ar_3p}, it enables us to
directly compare calculated results with experimental results.
In the following subsection, we demonstrate the comparison of theory with experiment.

\subsection{Comparison with experimental results}

\begin{figure}[htbp]
%\centering
\includegraphics[width=0.8\columnwidth]{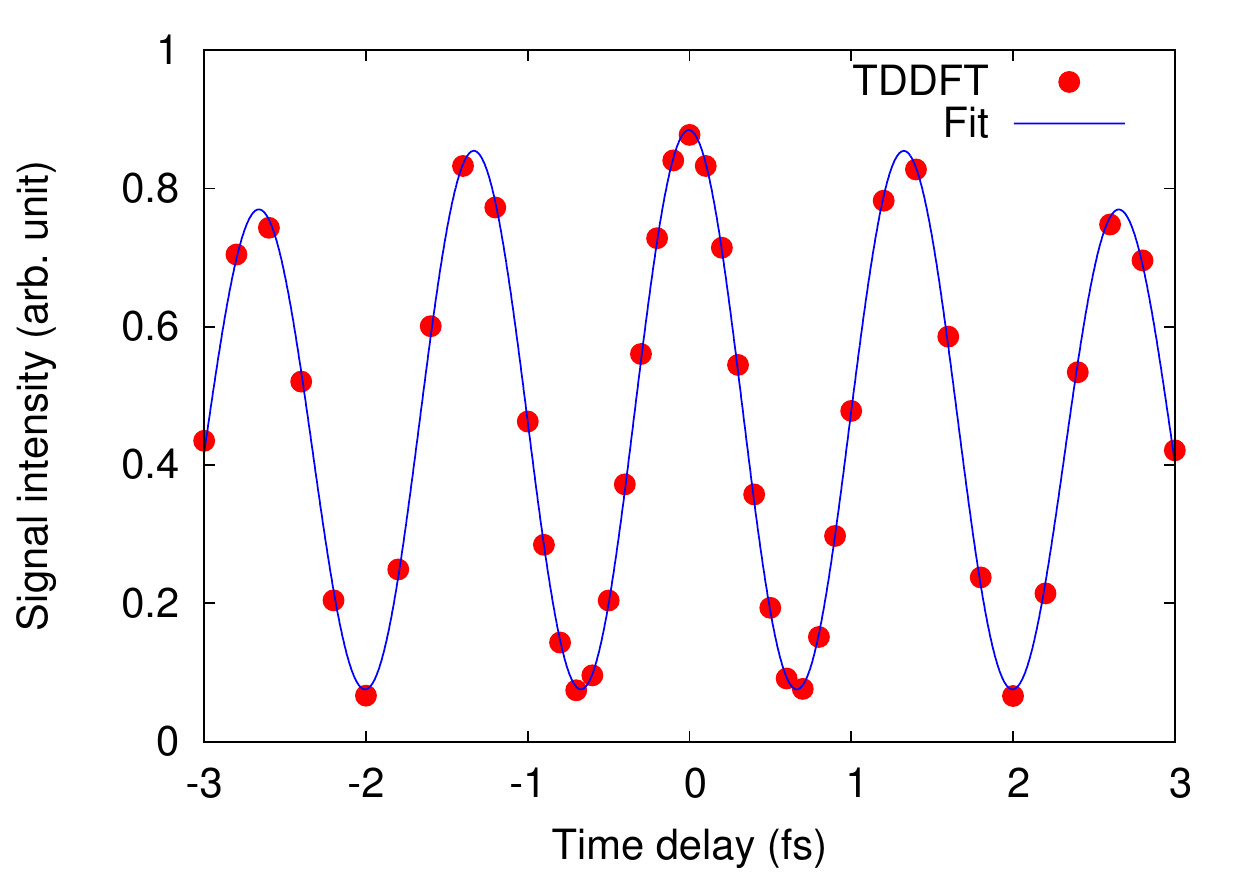}
\caption{\label{fig:Ar_3p_26th}
Extracted RABBITT trace of the Ar 3$p$ shell for the $26$th side-band. Red-points show the TDDFT result
for each delay. The blue line shows the fitting function in Eq. (\ref{eq:fitting}).
}
\end{figure}

Here, we compare the computed time delays from TDDFT simulations with the experimental results \cite{PhysRevLett.106.143002,PhysRevA.85.053424}.
For this purpose, we first numerically extract the delay from the RABBITT trace 
in Fig. \ref{fig:RABBITT_Ar_3p}.
To extract the delay for each side band, we average the RABBITT trace around the central frequency
of the side-band with width of $  \omega_0/2$. For example, to extract the 26th side band 
in Fig. \ref{fig:RABBITT_Ar_3p}, we average the signal between $26 \omega_0 \pm   \omega_0/4$.
Figure \ref{fig:Ar_3p_26th} shows the extracted signal for the 26th side-band 
in Fig. \ref{fig:RABBITT_Ar_3p}. Each red-point shows the result from
a single TDDFT simulation with the corresponding time delay.
In order to extract the time delay, we further fit the numerical signal by an analytic function of the following form:
\be
S(t)=A\cdot \cos^4 \left[ \frac{\pi}{\sigma} (t-t_0)\right] 
\cos^2 \left[\omega_0 (t-\tau_{delay}) \right ] +C,
\label{eq:fitting}
\ee
where $A$, $C$, $\sigma$, $t_0$, and $\tau_{delay}$ are fitting parameters. Here, $\tau_{delay}$ is 
the time delay, which we aim to extract.
In Fig. \ref{fig:Ar_3p_26th}, the fitting function is shown as a blue line.
One sees that the fitting function represents the signal very well over the entire time delay range.

Even though the absolute time delay with respect to the IR field can be readily extracted from these theoretical simulations,
it is in fact hard to extract this absolute delay from experimental results, since the absolute time-zero
cannot be determined in the experiments.
Therefore, so far, experimental results only provide the relative time delay between 
two different excitation channels such as excitation from different atomic shells.

\begin{figure}[htbp]
%\centering
\includegraphics[width=0.8\columnwidth]{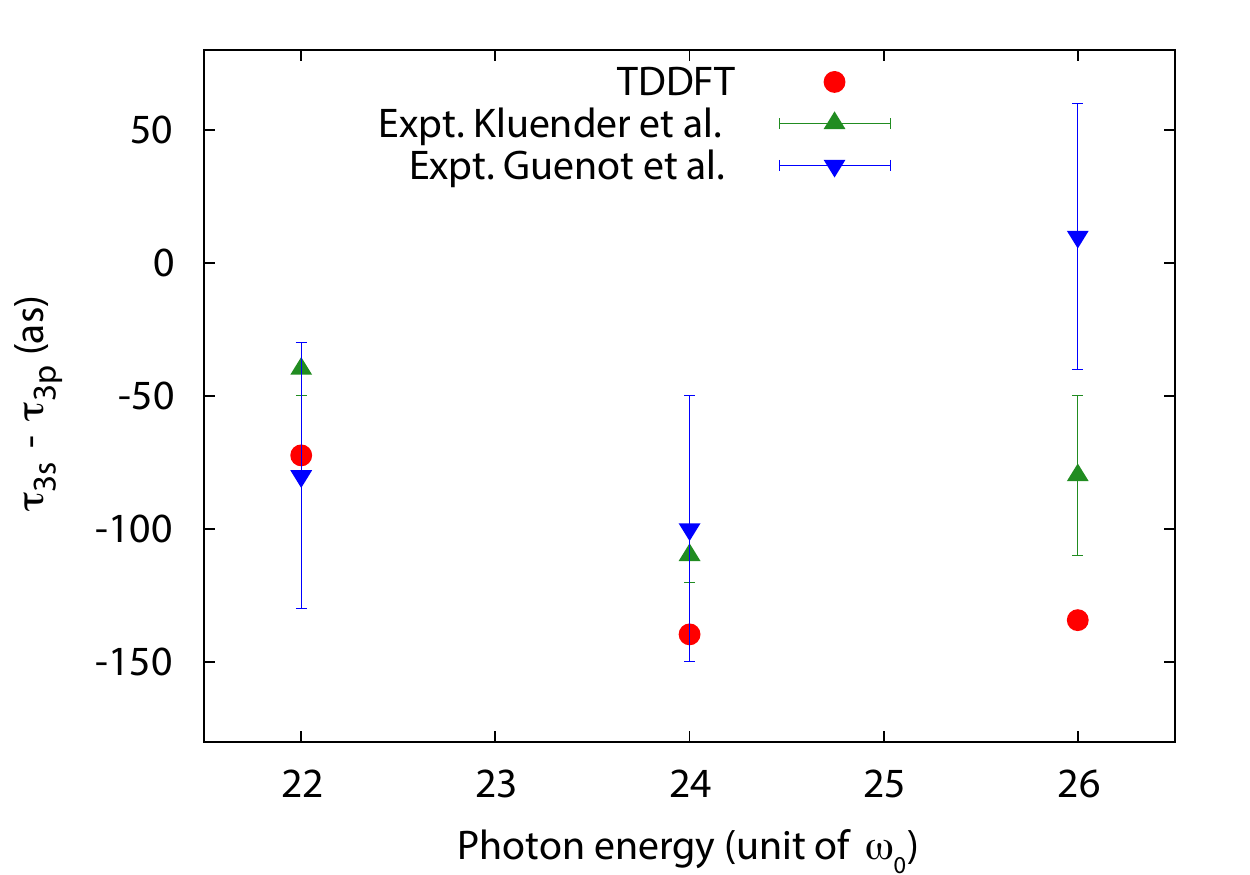}
\caption{\label{fig:delay_Ar_3s_3p}
Comparison of the delay differences for ionization of Ar from the 3s and 3p shells:
the theoretical results by the TDDFT with ALDA + ADSIC (red-circles), 
the experimental results by Kl\"under \textit{et al.}
\cite{PhysRevLett.106.143002} (green up-pointing triangles), and the experimental results by Gu\'enot \textit{et al} \cite{PhysRevA.85.053424}.
(blue down-pointing triangles) are shown.
}
\end{figure}

Figure \ref{fig:delay_Ar_3s_3p} shows the relative time delay for ionization of Ar from the 3$s$ and 3$p$
shells. Red-circles show the TDDFT results, while up and down-pointing triangles show 
recent experimental results \cite{PhysRevLett.106.143002,PhysRevA.85.053424}.
One sees that the TDDFT results are in excellent agreement with the experimental results.

We note that while our work based on the TDDFT with ALDA and ADSIC shows very good agreement with the experiment, 
recent results by Magrakvelidze \textit{et al.} \cite{PhysRevA.91.063415}, also based on TDDFT in the local density approximation,
appears to disagree on the same experiment.
Here, we discuss a possible origin for this apparent inconsistency and suggest that emerges from the separation of the photoemission delay into 
two consecutive steps -- an approach shared by many RABBITT models.
In many works that deal with modelling RABBITT \cite{PhysRevLett.106.143002,RevModPhys.87.765}, the time delay is first decomposed 
into two components:
\be
\tau_{delay} = \tau_{W} + \tau_{cc},
\ee
where $\tau_{W}$ is so-called Wigner delay due to the EUV single-photon 
absorption process \cite{PhysRev.98.145,PhysRev.118.349},
and $\tau_{cc}$ is the so-called continuum-continuum delay due to the additional contribution
from the IR field \cite{PhysRevLett.106.143002,DAHLSTROM201353}. These two delays are often treated and computed separately.
In contrast, in the present work, we do not rely on such a decomposition of the RABBITT delay,
but directly compute the total delay, by simulating the whole measurement processes, starting from two external laser fields 
and the system in its groundstate and by performing a time-propagation all the way to the detection of the emitted photo-electron.
As a result our method treats the excitation, emission and detection process on the same footing and, as shown in the present work, succeeds to accurately reproduce the experiment.
This fact indicates that a fully consistent treatment for all the delay components
is significant to correctly understand experimental results, and thus,
a direct simulation of the entire measurement processes is required.
Furthermore, a separated treatment of the delay components is highly non-trivial
for complex systems such as large molecules or solid-state surfaces.
In the last case, an additional delay related to the electron transport towards the surface has to be taken into account~\cite{Locher:15}.

\subsection{Many-body effects}

One of the strong points of the present TDDFT photoelectron spectroscopy is the capability to
investigate the impact of many-body effects directly in the experimental observable.
Therefore, the present TDDFT simulation of photoelectron spectroscopy offers novel opportunities to explore the role
of many-body effects in the photoelectron emission process.
To demonstrate this capability, we here investigate the role of the dynamical electron-electron interaction effects in the argon RABBITT spectroscopy. In our calculation we employ the local density approximation (LDA) for all exchange and correlation effects that are beyond the time-dependent Hartree approximation. While LDA is known to not be able to represent most exchange effects and only weak correlation, the dynamical Hartree potential we use in our calculation should be expected to have a large impact on the dynamics. Since, for the photo-emission process we require a functional with SIC-correction it is not possible to completely separate the effect of the xc-functional and the Hartree potential, but we posit that our results can be considered to be roughly equivalent to at least a random phase approximation level of theory.  

To demonstrate the influence of some of the many-body effects as captured by the current approximation, we additionally perform TDDFT RABBITT simulations, where we neglect the time-dependence of the Hartree and the exchange-correlation potentials. That is to say that throughout the propagation of the KS equation, the Kohn-Sham potential is kept "frozen" to the ground state.
This treatment corresponds to the independent particle (IP) approximation since
all the electrons independently move in a common and fixed mean-field potential.

\begin{figure}[htbp]
%\centering
\includegraphics[width=0.8\columnwidth]{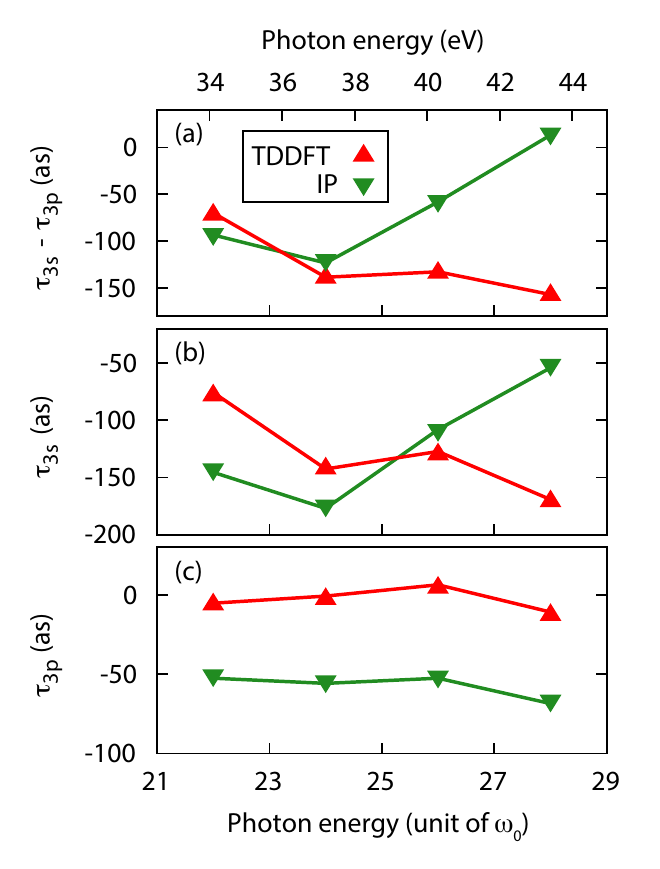}
\caption{\label{fig:delay_Ar_comp_tddft_ip}
Comparison of the RABBITT delay with and without the many-body effects as captured by TDDFT with LDA.
The panel (a) shows the relative Ar 3$s$-3$p$ delay. The individual delays of the Ar 3$s$ and  
3$p$ shells are also shown in the panels (b) and (c), respectively.
}
\end{figure}

Figure \ref{fig:delay_Ar_comp_tddft_ip} (a) shows the relative delays $\tau_{3s}-\tau_{3p}$ computed
by the TDDFT and the IP calculations.
One sees that, while the two calculations provide the similar relative delays
in the lower photon energy region, they show a discrepancy in the higher energy region.
In this high energy range around 42 eV, the Ar 3$s$ photoionization cross section becomes very small
due to many-electron effects \cite{PhysRevA.47.3888}. This region is the so-called Cooper minimum,
and the influence of the Cooper minima in the photoionization delay has been intensively discussed \cite{RevModPhys.87.765}.
Previous TDDFT calculations with this ADSIC reported 
a photoionization cross section in good agreement with 
the experiment~\cite{PhysRevA.90.033412}.

To obtain further insight of the impact of the many-body effects in the photoionization delay in atoms,
we investigate the RABBITT delay for individual Ar 3$s$ and 3$p$ shells.
Figure \ref{fig:delay_Ar_comp_tddft_ip} (b) and (c) show the RABBITT delays for 
Ar 3$s$ and 3$p$ shells, respectively.
As seen from Fig. \ref{fig:delay_Ar_comp_tddft_ip} (b), many-body effects
play different roles for the photoionization from Ar 3$s$ shell in the lower and higher energy ranges: 
while the many-body interaction induces a positive delay in the lower energy range, it induces 
a negative delay in the higher energy range. 
This fact indicates a correlation among many-body effects, Cooper minima, and the photoionization delay.
In contrast, as seen from Fig. \ref{fig:delay_Ar_comp_tddft_ip} (c), the many-body effects
uniformly increase the delay of the Ar 3$p$ shell in all the investigated photon energy range.
Importantly, one sees that it induces similar amount of positive delay for both
Ar 3$s$ and 3$p$ shells in the low photon energy range. Therefore, the influence of 
many-body effects on the relative 3$s$-3$p$ delay in the low photon energy range 
is cancelled out and appears to have no influence on the relative delay.

%============================================================================================================
\section{Summary \label{sec:summary}}
%============================================================================================================

In this work, we developed an efficient first-principles attosecond photoelectron spectroscopy technique
based on time-dependent density functional theory (TDDFT),
focusing on the reconstruction of attosecond beating by interference of two-photon transition (RABBITT).
We applied the TDDFT RABBITT simulation to investigate the photoemission from the 3$s$ and 3$p$ shells of Argon.
We demonstrated that the TDDFT results nicely reproduce recent experimental results \cite{PhysRevLett.106.143002,PhysRevA.85.053424}.
The good agreement of our TDDFT simulation with the experimental results is apparent inconsistency with 
previous work that also employs TDDFT \cite{PhysRevA.91.063415}; Magrakvelidze, \textit{et al.} reported
that the results computed by TDDFT with the local density approximation disagrees with 
the measured relative Ar 3$s$-3$p$ time delay.
While the previous work computed only the Wigner delay with TDDFT but employed another 
theory to treat the continuum-continuum delay, the present work treats all the components of 
the delay at the same level in the TDDFT propagation. Therefore, the apparent inconsistency between the present and previous
works may originate from the inconsistent treatment of individual delay contributions
of the previous work. This fact indicates the significance of 
a consistent treatment for each delay contribution and the direct simulation of 
the whole measurement processes.
Furthermore, once target systems become complex, such as large molecules and solid-state
surfaces, this kind of step-wise approach to the complete delay becomes nontrivial or unfeasible.
Therefore, the fully consistent simulations for the whole measurement processes naturally emerges as 
a significant tool to attain microscopic insight of such attosecond experiments.

Furthermore, the presented TDDFT approach offers novel opportunities to investigate the role of
microscopic many-body effects in the photoemission process. In this work we have shown how, by freezing 
the time-dependent Hartree and exchange-correlation potentials, the role of many-body interactions
can be systematically investigated.

As a result, it turned out that many-body effects substantially affect the RABBITT 
photoionization delay.
In particular, we found that the induced delay in Ar 3$s$ photoionization changes its sign around the Cooper minimum. 
At the moment, accurate description of the exchange-correlation potential as well as 
electron-ion coupling is limited, and thus, many-body effects are not fully captured by our  
TDDFT simulation. However, once a better description for electron-electron and electron-ion
interactions is developed, the TDDFT RABBITT simulation could be employed to investigate
the role of decoherence due to electron-electron, electron-ion and electron-phonon
scattering both in the photoemission as well as the transport processes.

While in this work we presented results on a simple system such as gas-phase Argon, 
the current technique can be readily employed to more complex targets. 
In particular, the current approach as well as our implementation can be already used to investigate attosecond photoelectron dynamics of solid surfaces. 
It therefore represents a very powerful and timely technique to guide state-of-the-art experiments, and indeed, 
work along these lines with experiments is already underway.

\section*{Acknowledgements}
We thank L. Gallmann for carefully reading the manuscript and providing valuable comments. We thank U. Keller for helpful discussions and insight into this problem.
We acknowledge financial support from the European Research Council (ERC-2015-AdG-694097), Grupos Consolidados (IT578-13) and the European Union’s Horizon 2020 Research and Innovation program under Grant Agreements no. 676580 (NOMAD). S. A. S. acknowledges support by 
Alexander von Humboldt Foundation.

% \bibliography{rabbitt}

\begin{thebibliography}{55}%
\makeatletter
\providecommand \@ifxundefined [1]{%
 \@ifx{#1\undefined}
}%
\providecommand \@ifnum [1]{%
 \ifnum #1\expandafter \@firstoftwo
 \else \expandafter \@secondoftwo
 \fi
}%
\providecommand \@ifx [1]{%
 \ifx #1\expandafter \@firstoftwo
 \else \expandafter \@secondoftwo
 \fi
}%
\providecommand \natexlab [1]{#1}%
\providecommand \enquote  [1]{``#1''}%
\providecommand \bibnamefont  [1]{#1}%
\providecommand \bibfnamefont [1]{#1}%
\providecommand \citenamefont [1]{#1}%
\providecommand \href@noop [0]{\@secondoftwo}%
\providecommand \href [0]{\begingroup \@sanitize@url \@href}%
\providecommand \@href[1]{\@@startlink{#1}\@@href}%
\providecommand \@@href[1]{\endgroup#1\@@endlink}%
\providecommand \@sanitize@url [0]{\catcode `\\12\catcode `\$12\catcode
  `\&12\catcode `\#12\catcode `\^12\catcode `\_12\catcode `\%12\relax}%
\providecommand \@@startlink[1]{}%
\providecommand \@@endlink[0]{}%
\providecommand \url  [0]{\begingroup\@sanitize@url \@url }%
\providecommand \@url [1]{\endgroup\@href {#1}{\urlprefix }}%
\providecommand \urlprefix  [0]{URL }%
\providecommand \Eprint [0]{\href }%
\providecommand \doibase [0]{http://dx.doi.org/}%
\providecommand \selectlanguage [0]{\@gobble}%
\providecommand \bibinfo  [0]{\@secondoftwo}%
\providecommand \bibfield  [0]{\@secondoftwo}%
\providecommand \translation [1]{[#1]}%
\providecommand \BibitemOpen [0]{}%
\providecommand \bibitemStop [0]{}%
\providecommand \bibitemNoStop [0]{.\EOS\space}%
\providecommand \EOS [0]{\spacefactor3000\relax}%
\providecommand \BibitemShut  [1]{\csname bibitem#1\endcsname}%
\let\auto@bib@innerbib\@empty
%</preamble>
\bibitem [{\citenamefont {Hentschel}\ \emph {et~al.}(2001)\citenamefont
  {Hentschel}, \citenamefont {Kienberger}, \citenamefont {Spielmann},
  \citenamefont {Reider}, \citenamefont {Milosevic}, \citenamefont {Brabec},
  \citenamefont {Corkum}, \citenamefont {Heinzmann}, \citenamefont {Drescher},\
  and\ \citenamefont {Krausz}}]{hentschel2001attosecond}%
  \BibitemOpen
  \bibfield  {author} {\bibinfo {author} {\bibfnamefont {M.}~\bibnamefont
  {Hentschel}}, \bibinfo {author} {\bibfnamefont {R.}~\bibnamefont
  {Kienberger}}, \bibinfo {author} {\bibfnamefont {C.}~\bibnamefont
  {Spielmann}}, \bibinfo {author} {\bibfnamefont {G.~A.}\ \bibnamefont
  {Reider}}, \bibinfo {author} {\bibfnamefont {N.}~\bibnamefont {Milosevic}},
  \bibinfo {author} {\bibfnamefont {T.}~\bibnamefont {Brabec}}, \bibinfo
  {author} {\bibfnamefont {P.}~\bibnamefont {Corkum}}, \bibinfo {author}
  {\bibfnamefont {U.}~\bibnamefont {Heinzmann}}, \bibinfo {author}
  {\bibfnamefont {M.}~\bibnamefont {Drescher}}, \ and\ \bibinfo {author}
  {\bibfnamefont {F.}~\bibnamefont {Krausz}},\ }\href@noop {} {\bibfield
  {journal} {\bibinfo  {journal} {Nature}\ }\textbf {\bibinfo {volume} {414}},\
  \bibinfo {pages} {509} (\bibinfo {year} {2001})}\BibitemShut {NoStop}%
\bibitem [{\citenamefont {Krausz}\ and\ \citenamefont
  {Ivanov}(2009)}]{RevModPhys.81.163}%
  \BibitemOpen
  \bibfield  {author} {\bibinfo {author} {\bibfnamefont {F.}~\bibnamefont
  {Krausz}}\ and\ \bibinfo {author} {\bibfnamefont {M.}~\bibnamefont
  {Ivanov}},\ }\href@noop {} {\bibfield  {journal} {\bibinfo  {journal} {Rev.
  Mod. Phys.}\ }\textbf {\bibinfo {volume} {81}},\ \bibinfo {pages} {163}
  (\bibinfo {year} {2009})}\BibitemShut {NoStop}%
\bibitem [{\citenamefont {Krausz}\ and\ \citenamefont
  {Stockman}(2014)}]{krausz2014attosecond}%
  \BibitemOpen
  \bibfield  {author} {\bibinfo {author} {\bibfnamefont {F.}~\bibnamefont
  {Krausz}}\ and\ \bibinfo {author} {\bibfnamefont {M.~I.}\ \bibnamefont
  {Stockman}},\ }\href@noop {} {\bibfield  {journal} {\bibinfo  {journal}
  {Nature Photonics}\ }\textbf {\bibinfo {volume} {8}},\ \bibinfo {pages} {205}
  (\bibinfo {year} {2014})}\BibitemShut {NoStop}%
\bibitem [{\citenamefont {Goulielmakis}\ \emph {et~al.}(2010)\citenamefont
  {Goulielmakis}, \citenamefont {Loh}, \citenamefont {Wirth}, \citenamefont
  {Santra}, \citenamefont {Rohringer}, \citenamefont {Yakovlev}, \citenamefont
  {Zherebtsov}, \citenamefont {Pfeifer}, \citenamefont {Azzeer}, \citenamefont
  {Kling} \emph {et~al.}}]{goulielmakis2010real}%
  \BibitemOpen
  \bibfield  {author} {\bibinfo {author} {\bibfnamefont {E.}~\bibnamefont
  {Goulielmakis}}, \bibinfo {author} {\bibfnamefont {Z.-H.}\ \bibnamefont
  {Loh}}, \bibinfo {author} {\bibfnamefont {A.}~\bibnamefont {Wirth}}, \bibinfo
  {author} {\bibfnamefont {R.}~\bibnamefont {Santra}}, \bibinfo {author}
  {\bibfnamefont {N.}~\bibnamefont {Rohringer}}, \bibinfo {author}
  {\bibfnamefont {V.~S.}\ \bibnamefont {Yakovlev}}, \bibinfo {author}
  {\bibfnamefont {S.}~\bibnamefont {Zherebtsov}}, \bibinfo {author}
  {\bibfnamefont {T.}~\bibnamefont {Pfeifer}}, \bibinfo {author} {\bibfnamefont
  {A.~M.}\ \bibnamefont {Azzeer}}, \bibinfo {author} {\bibfnamefont {M.~F.}\
  \bibnamefont {Kling}},  \emph {et~al.},\ }\href@noop {} {\bibfield  {journal}
  {\bibinfo  {journal} {Nature}\ }\textbf {\bibinfo {volume} {466}},\ \bibinfo
  {pages} {739} (\bibinfo {year} {2010})}\BibitemShut {NoStop}%
\bibitem [{\citenamefont {Wang}\ \emph {et~al.}(2010)\citenamefont {Wang},
  \citenamefont {Chini}, \citenamefont {Chen}, \citenamefont {Zhang},
  \citenamefont {He}, \citenamefont {Cheng}, \citenamefont {Wu}, \citenamefont
  {Thumm},\ and\ \citenamefont {Chang}}]{PhysRevLett.105.143002}%
  \BibitemOpen
  \bibfield  {author} {\bibinfo {author} {\bibfnamefont {H.}~\bibnamefont
  {Wang}}, \bibinfo {author} {\bibfnamefont {M.}~\bibnamefont {Chini}},
  \bibinfo {author} {\bibfnamefont {S.}~\bibnamefont {Chen}}, \bibinfo {author}
  {\bibfnamefont {C.-H.}\ \bibnamefont {Zhang}}, \bibinfo {author}
  {\bibfnamefont {F.}~\bibnamefont {He}}, \bibinfo {author} {\bibfnamefont
  {Y.}~\bibnamefont {Cheng}}, \bibinfo {author} {\bibfnamefont
  {Y.}~\bibnamefont {Wu}}, \bibinfo {author} {\bibfnamefont {U.}~\bibnamefont
  {Thumm}}, \ and\ \bibinfo {author} {\bibfnamefont {Z.}~\bibnamefont
  {Chang}},\ }\href@noop {} {\bibfield  {journal} {\bibinfo  {journal} {Phys.
  Rev. Lett.}\ }\textbf {\bibinfo {volume} {105}},\ \bibinfo {pages} {143002}
  (\bibinfo {year} {2010})}\BibitemShut {NoStop}%
\bibitem [{\citenamefont {Holler}\ \emph {et~al.}(2011)\citenamefont {Holler},
  \citenamefont {Schapper}, \citenamefont {Gallmann},\ and\ \citenamefont
  {Keller}}]{PhysRevLett.106.123601}%
  \BibitemOpen
  \bibfield  {author} {\bibinfo {author} {\bibfnamefont {M.}~\bibnamefont
  {Holler}}, \bibinfo {author} {\bibfnamefont {F.}~\bibnamefont {Schapper}},
  \bibinfo {author} {\bibfnamefont {L.}~\bibnamefont {Gallmann}}, \ and\
  \bibinfo {author} {\bibfnamefont {U.}~\bibnamefont {Keller}},\ }\href@noop {}
  {\bibfield  {journal} {\bibinfo  {journal} {Phys. Rev. Lett.}\ }\textbf
  {\bibinfo {volume} {106}},\ \bibinfo {pages} {123601} (\bibinfo {year}
  {2011})}\BibitemShut {NoStop}%
\bibitem [{\citenamefont {Paul}\ \emph {et~al.}(2001)\citenamefont {Paul},
  \citenamefont {Toma}, \citenamefont {Breger}, \citenamefont {Mullot},
  \citenamefont {Aug{\'e}}, \citenamefont {Balcou}, \citenamefont {Muller},\
  and\ \citenamefont {Agostini}}]{Paul1689}%
  \BibitemOpen
  \bibfield  {author} {\bibinfo {author} {\bibfnamefont {P.~M.}\ \bibnamefont
  {Paul}}, \bibinfo {author} {\bibfnamefont {E.~S.}\ \bibnamefont {Toma}},
  \bibinfo {author} {\bibfnamefont {P.}~\bibnamefont {Breger}}, \bibinfo
  {author} {\bibfnamefont {G.}~\bibnamefont {Mullot}}, \bibinfo {author}
  {\bibfnamefont {F.}~\bibnamefont {Aug{\'e}}}, \bibinfo {author}
  {\bibfnamefont {P.}~\bibnamefont {Balcou}}, \bibinfo {author} {\bibfnamefont
  {H.~G.}\ \bibnamefont {Muller}}, \ and\ \bibinfo {author} {\bibfnamefont
  {P.}~\bibnamefont {Agostini}},\ }\href@noop {} {\bibfield  {journal}
  {\bibinfo  {journal} {Science}\ }\textbf {\bibinfo {volume} {292}},\ \bibinfo
  {pages} {1689} (\bibinfo {year} {2001})}\BibitemShut {NoStop}%
\bibitem [{\citenamefont {Muller}(2002)}]{muller2002reconstruction}%
  \BibitemOpen
  \bibfield  {author} {\bibinfo {author} {\bibfnamefont {H.~.~G.}\ \bibnamefont
  {Muller}},\ }\href@noop {} {\bibfield  {journal} {\bibinfo  {journal}
  {Applied Physics B}\ }\textbf {\bibinfo {volume} {74}},\ \bibinfo {pages}
  {s17} (\bibinfo {year} {2002})}\BibitemShut {NoStop}%
\bibitem [{\citenamefont {Itatani}\ \emph {et~al.}(2002)\citenamefont
  {Itatani}, \citenamefont {Qu\'er\'e}, \citenamefont {Yudin}, \citenamefont
  {Ivanov}, \citenamefont {Krausz},\ and\ \citenamefont
  {Corkum}}]{PhysRevLett.88.173903}%
  \BibitemOpen
  \bibfield  {author} {\bibinfo {author} {\bibfnamefont {J.}~\bibnamefont
  {Itatani}}, \bibinfo {author} {\bibfnamefont {F.}~\bibnamefont {Qu\'er\'e}},
  \bibinfo {author} {\bibfnamefont {G.~L.}\ \bibnamefont {Yudin}}, \bibinfo
  {author} {\bibfnamefont {M.~Y.}\ \bibnamefont {Ivanov}}, \bibinfo {author}
  {\bibfnamefont {F.}~\bibnamefont {Krausz}}, \ and\ \bibinfo {author}
  {\bibfnamefont {P.~B.}\ \bibnamefont {Corkum}},\ }\href@noop {} {\bibfield
  {journal} {\bibinfo  {journal} {Phys. Rev. Lett.}\ }\textbf {\bibinfo
  {volume} {88}},\ \bibinfo {pages} {173903} (\bibinfo {year}
  {2002})}\BibitemShut {NoStop}%
\bibitem [{\citenamefont {Kienberger}\ \emph {et~al.}(2004)\citenamefont
  {Kienberger}, \citenamefont {Goulielmakis}, \citenamefont {Uiberacker},
  \citenamefont {Baltuska}, \citenamefont {Yakovlev}, \citenamefont {Bammer},
  \citenamefont {Scrinzi}, \citenamefont {Westerwalbesloh}, \citenamefont
  {Kleineberg}, \citenamefont {Heinzmann} \emph
  {et~al.}}]{kienberger2004atomic}%
  \BibitemOpen
  \bibfield  {author} {\bibinfo {author} {\bibfnamefont {R.}~\bibnamefont
  {Kienberger}}, \bibinfo {author} {\bibfnamefont {E.}~\bibnamefont
  {Goulielmakis}}, \bibinfo {author} {\bibfnamefont {M.}~\bibnamefont
  {Uiberacker}}, \bibinfo {author} {\bibfnamefont {A.}~\bibnamefont
  {Baltuska}}, \bibinfo {author} {\bibfnamefont {V.}~\bibnamefont {Yakovlev}},
  \bibinfo {author} {\bibfnamefont {F.}~\bibnamefont {Bammer}}, \bibinfo
  {author} {\bibfnamefont {A.}~\bibnamefont {Scrinzi}}, \bibinfo {author}
  {\bibfnamefont {T.}~\bibnamefont {Westerwalbesloh}}, \bibinfo {author}
  {\bibfnamefont {U.}~\bibnamefont {Kleineberg}}, \bibinfo {author}
  {\bibfnamefont {U.}~\bibnamefont {Heinzmann}},  \emph {et~al.},\ }\href@noop
  {} {\bibfield  {journal} {\bibinfo  {journal} {Nature}\ }\textbf {\bibinfo
  {volume} {427}},\ \bibinfo {pages} {817} (\bibinfo {year}
  {2004})}\BibitemShut {NoStop}%
\bibitem [{\citenamefont {Beck}\ \emph {et~al.}(2014)\citenamefont {Beck},
  \citenamefont {Bernhardt}, \citenamefont {Warrick}, \citenamefont {Wu},
  \citenamefont {Chen}, \citenamefont {Gaarde}, \citenamefont {Schafer},
  \citenamefont {Neumark},\ and\ \citenamefont {Leone}}]{beck2014attosecond}%
  \BibitemOpen
  \bibfield  {author} {\bibinfo {author} {\bibfnamefont {A.~R.}\ \bibnamefont
  {Beck}}, \bibinfo {author} {\bibfnamefont {B.}~\bibnamefont {Bernhardt}},
  \bibinfo {author} {\bibfnamefont {E.~R.}\ \bibnamefont {Warrick}}, \bibinfo
  {author} {\bibfnamefont {M.}~\bibnamefont {Wu}}, \bibinfo {author}
  {\bibfnamefont {S.}~\bibnamefont {Chen}}, \bibinfo {author} {\bibfnamefont
  {M.~B.}\ \bibnamefont {Gaarde}}, \bibinfo {author} {\bibfnamefont {K.~J.}\
  \bibnamefont {Schafer}}, \bibinfo {author} {\bibfnamefont {D.~M.}\
  \bibnamefont {Neumark}}, \ and\ \bibinfo {author} {\bibfnamefont {S.~R.}\
  \bibnamefont {Leone}},\ }\href@noop {} {\bibfield  {journal} {\bibinfo
  {journal} {New Journal of Physics}\ }\textbf {\bibinfo {volume} {16}},\
  \bibinfo {pages} {113016} (\bibinfo {year} {2014})}\BibitemShut {NoStop}%
\bibitem [{\citenamefont {Warrick}\ \emph {et~al.}(2016)\citenamefont
  {Warrick}, \citenamefont {Cao}, \citenamefont {Neumark},\ and\ \citenamefont
  {Leone}}]{warrick2016probing}%
  \BibitemOpen
  \bibfield  {author} {\bibinfo {author} {\bibfnamefont {E.~R.}\ \bibnamefont
  {Warrick}}, \bibinfo {author} {\bibfnamefont {W.}~\bibnamefont {Cao}},
  \bibinfo {author} {\bibfnamefont {D.~M.}\ \bibnamefont {Neumark}}, \ and\
  \bibinfo {author} {\bibfnamefont {S.~R.}\ \bibnamefont {Leone}},\ }\href@noop
  {} {\bibfield  {journal} {\bibinfo  {journal} {The Journal of Physical
  Chemistry A}\ }\textbf {\bibinfo {volume} {120}},\ \bibinfo {pages} {3165}
  (\bibinfo {year} {2016})}\BibitemShut {NoStop}%
\bibitem [{\citenamefont {Reduzzi}\ \emph {et~al.}(2016)\citenamefont
  {Reduzzi}, \citenamefont {Chu}, \citenamefont {Feng}, \citenamefont
  {Dubrouil}, \citenamefont {Hummert}, \citenamefont {Calegari}, \citenamefont
  {Frassetto}, \citenamefont {Poletto}, \citenamefont {Kornilov}, \citenamefont
  {Nisoli} \emph {et~al.}}]{reduzzi2016observation}%
  \BibitemOpen
  \bibfield  {author} {\bibinfo {author} {\bibfnamefont {M.}~\bibnamefont
  {Reduzzi}}, \bibinfo {author} {\bibfnamefont {W.}~\bibnamefont {Chu}},
  \bibinfo {author} {\bibfnamefont {C.}~\bibnamefont {Feng}}, \bibinfo {author}
  {\bibfnamefont {A.}~\bibnamefont {Dubrouil}}, \bibinfo {author}
  {\bibfnamefont {J.}~\bibnamefont {Hummert}}, \bibinfo {author} {\bibfnamefont
  {F.}~\bibnamefont {Calegari}}, \bibinfo {author} {\bibfnamefont
  {F.}~\bibnamefont {Frassetto}}, \bibinfo {author} {\bibfnamefont
  {L.}~\bibnamefont {Poletto}}, \bibinfo {author} {\bibfnamefont
  {O.}~\bibnamefont {Kornilov}}, \bibinfo {author} {\bibfnamefont
  {M.}~\bibnamefont {Nisoli}},  \emph {et~al.},\ }\href@noop {} {\bibfield
  {journal} {\bibinfo  {journal} {Journal of Physics B: Atomic, Molecular and
  Optical Physics}\ }\textbf {\bibinfo {volume} {49}},\ \bibinfo {pages}
  {065102} (\bibinfo {year} {2016})}\BibitemShut {NoStop}%
\bibitem [{\citenamefont {Santra}\ \emph {et~al.}(2011)\citenamefont {Santra},
  \citenamefont {Yakovlev}, \citenamefont {Pfeifer},\ and\ \citenamefont
  {Loh}}]{PhysRevA.83.033405}%
  \BibitemOpen
  \bibfield  {author} {\bibinfo {author} {\bibfnamefont {R.}~\bibnamefont
  {Santra}}, \bibinfo {author} {\bibfnamefont {V.~S.}\ \bibnamefont
  {Yakovlev}}, \bibinfo {author} {\bibfnamefont {T.}~\bibnamefont {Pfeifer}}, \
  and\ \bibinfo {author} {\bibfnamefont {Z.-H.}\ \bibnamefont {Loh}},\
  }\href@noop {} {\bibfield  {journal} {\bibinfo  {journal} {Phys. Rev. A}\
  }\textbf {\bibinfo {volume} {83}},\ \bibinfo {pages} {033405} (\bibinfo
  {year} {2011})}\BibitemShut {NoStop}%
\bibitem [{\citenamefont {De~Giovannini}\ \emph {et~al.}(2013)\citenamefont
  {De~Giovannini}, \citenamefont {Brunetto}, \citenamefont {Castro},
  \citenamefont {Walkenhorst},\ and\ \citenamefont
  {Rubio}}]{CPHC:CPHC201201007}%
  \BibitemOpen
  \bibfield  {author} {\bibinfo {author} {\bibfnamefont {U.}~\bibnamefont
  {De~Giovannini}}, \bibinfo {author} {\bibfnamefont {G.}~\bibnamefont
  {Brunetto}}, \bibinfo {author} {\bibfnamefont {A.}~\bibnamefont {Castro}},
  \bibinfo {author} {\bibfnamefont {J.}~\bibnamefont {Walkenhorst}}, \ and\
  \bibinfo {author} {\bibfnamefont {A.}~\bibnamefont {Rubio}},\ }\href@noop {}
  {\bibfield  {journal} {\bibinfo  {journal} {ChemPhysChem}\ }\textbf {\bibinfo
  {volume} {14}},\ \bibinfo {pages} {1363} (\bibinfo {year}
  {2013})}\BibitemShut {NoStop}%
\bibitem [{\citenamefont {Schultze}\ \emph {et~al.}(2014)\citenamefont
  {Schultze}, \citenamefont {Ramasesha}, \citenamefont {Pemmaraju},
  \citenamefont {Sato}, \citenamefont {Whitmore}, \citenamefont {Gandman},
  \citenamefont {Prell}, \citenamefont {Borja}, \citenamefont {Prendergast},
  \citenamefont {Yabana}, \citenamefont {Neumark},\ and\ \citenamefont
  {Leone}}]{Schultze1348}%
  \BibitemOpen
  \bibfield  {author} {\bibinfo {author} {\bibfnamefont {M.}~\bibnamefont
  {Schultze}}, \bibinfo {author} {\bibfnamefont {K.}~\bibnamefont {Ramasesha}},
  \bibinfo {author} {\bibfnamefont {C.}~\bibnamefont {Pemmaraju}}, \bibinfo
  {author} {\bibfnamefont {S.}~\bibnamefont {Sato}}, \bibinfo {author}
  {\bibfnamefont {D.}~\bibnamefont {Whitmore}}, \bibinfo {author}
  {\bibfnamefont {A.}~\bibnamefont {Gandman}}, \bibinfo {author} {\bibfnamefont
  {J.~S.}\ \bibnamefont {Prell}}, \bibinfo {author} {\bibfnamefont {L.~J.}\
  \bibnamefont {Borja}}, \bibinfo {author} {\bibfnamefont {D.}~\bibnamefont
  {Prendergast}}, \bibinfo {author} {\bibfnamefont {K.}~\bibnamefont {Yabana}},
  \bibinfo {author} {\bibfnamefont {D.~M.}\ \bibnamefont {Neumark}}, \ and\
  \bibinfo {author} {\bibfnamefont {S.~R.}\ \bibnamefont {Leone}},\ }\href@noop
  {} {\bibfield  {journal} {\bibinfo  {journal} {Science}\ }\textbf {\bibinfo
  {volume} {346}},\ \bibinfo {pages} {1348} (\bibinfo {year}
  {2014})}\BibitemShut {NoStop}%
\bibitem [{\citenamefont {Lucchini}\ \emph {et~al.}(2016)\citenamefont
  {Lucchini}, \citenamefont {Sato}, \citenamefont {Ludwig}, \citenamefont
  {Herrmann}, \citenamefont {Volkov}, \citenamefont {Kasmi}, \citenamefont
  {Shinohara}, \citenamefont {Yabana}, \citenamefont {Gallmann},\ and\
  \citenamefont {Keller}}]{Lucchini916}%
  \BibitemOpen
  \bibfield  {author} {\bibinfo {author} {\bibfnamefont {M.}~\bibnamefont
  {Lucchini}}, \bibinfo {author} {\bibfnamefont {S.~A.}\ \bibnamefont {Sato}},
  \bibinfo {author} {\bibfnamefont {A.}~\bibnamefont {Ludwig}}, \bibinfo
  {author} {\bibfnamefont {J.}~\bibnamefont {Herrmann}}, \bibinfo {author}
  {\bibfnamefont {M.}~\bibnamefont {Volkov}}, \bibinfo {author} {\bibfnamefont
  {L.}~\bibnamefont {Kasmi}}, \bibinfo {author} {\bibfnamefont
  {Y.}~\bibnamefont {Shinohara}}, \bibinfo {author} {\bibfnamefont
  {K.}~\bibnamefont {Yabana}}, \bibinfo {author} {\bibfnamefont
  {L.}~\bibnamefont {Gallmann}}, \ and\ \bibinfo {author} {\bibfnamefont
  {U.}~\bibnamefont {Keller}},\ }\href@noop {} {\bibfield  {journal} {\bibinfo
  {journal} {Science}\ }\textbf {\bibinfo {volume} {353}},\ \bibinfo {pages}
  {916} (\bibinfo {year} {2016})}\BibitemShut {NoStop}%
\bibitem [{\citenamefont {Z{\"u}rch}\ \emph {et~al.}(2017)\citenamefont
  {Z{\"u}rch}, \citenamefont {Chang}, \citenamefont {Borja}, \citenamefont
  {Kraus}, \citenamefont {Cushing}, \citenamefont {Gandman}, \citenamefont
  {Kaplan}, \citenamefont {Oh}, \citenamefont {Prell}, \citenamefont
  {Prendergast}, \citenamefont {Pemmaraju}, \citenamefont {Neumark},\ and\
  \citenamefont {Leone}}]{zurch2017direct}%
  \BibitemOpen
  \bibfield  {author} {\bibinfo {author} {\bibfnamefont {M.}~\bibnamefont
  {Z{\"u}rch}}, \bibinfo {author} {\bibfnamefont {H.-T.}\ \bibnamefont
  {Chang}}, \bibinfo {author} {\bibfnamefont {L.~J.}\ \bibnamefont {Borja}},
  \bibinfo {author} {\bibfnamefont {P.~M.}\ \bibnamefont {Kraus}}, \bibinfo
  {author} {\bibfnamefont {S.~K.}\ \bibnamefont {Cushing}}, \bibinfo {author}
  {\bibfnamefont {A.}~\bibnamefont {Gandman}}, \bibinfo {author} {\bibfnamefont
  {C.~J.}\ \bibnamefont {Kaplan}}, \bibinfo {author} {\bibfnamefont {M.~H.}\
  \bibnamefont {Oh}}, \bibinfo {author} {\bibfnamefont {J.~S.}\ \bibnamefont
  {Prell}}, \bibinfo {author} {\bibfnamefont {D.}~\bibnamefont {Prendergast}},
  \bibinfo {author} {\bibfnamefont {C.~D.}\ \bibnamefont {Pemmaraju}}, \bibinfo
  {author} {\bibfnamefont {D.~M.}\ \bibnamefont {Neumark}}, \ and\ \bibinfo
  {author} {\bibfnamefont {S.~R.}\ \bibnamefont {Leone}},\ }\href@noop {}
  {\bibfield  {journal} {\bibinfo  {journal} {Nature communications}\ }\textbf
  {\bibinfo {volume} {8}},\ \bibinfo {pages} {15734} (\bibinfo {year}
  {2017})}\BibitemShut {NoStop}%
\bibitem [{\citenamefont {Hohenberg}\ and\ \citenamefont
  {Kohn}(1964)}]{PhysRev.136.B864}%
  \BibitemOpen
  \bibfield  {author} {\bibinfo {author} {\bibfnamefont {P.}~\bibnamefont
  {Hohenberg}}\ and\ \bibinfo {author} {\bibfnamefont {W.}~\bibnamefont
  {Kohn}},\ }\href@noop {} {\bibfield  {journal} {\bibinfo  {journal} {Phys.
  Rev.}\ }\textbf {\bibinfo {volume} {136}},\ \bibinfo {pages} {B864} (\bibinfo
  {year} {1964})}\BibitemShut {NoStop}%
\bibitem [{\citenamefont {Kohn}\ and\ \citenamefont
  {Sham}(1965)}]{PhysRev.140.A1133}%
  \BibitemOpen
  \bibfield  {author} {\bibinfo {author} {\bibfnamefont {W.}~\bibnamefont
  {Kohn}}\ and\ \bibinfo {author} {\bibfnamefont {L.~J.}\ \bibnamefont
  {Sham}},\ }\href@noop {} {\bibfield  {journal} {\bibinfo  {journal} {Phys.
  Rev.}\ }\textbf {\bibinfo {volume} {140}},\ \bibinfo {pages} {A1133}
  (\bibinfo {year} {1965})}\BibitemShut {NoStop}%
\bibitem [{\citenamefont {Runge}\ and\ \citenamefont
  {Gross}(1984)}]{PhysRevLett.52.997}%
  \BibitemOpen
  \bibfield  {author} {\bibinfo {author} {\bibfnamefont {E.}~\bibnamefont
  {Runge}}\ and\ \bibinfo {author} {\bibfnamefont {E.~K.~U.}\ \bibnamefont
  {Gross}},\ }\href@noop {} {\bibfield  {journal} {\bibinfo  {journal} {Phys.
  Rev. Lett.}\ }\textbf {\bibinfo {volume} {52}},\ \bibinfo {pages} {997}
  (\bibinfo {year} {1984})}\BibitemShut {NoStop}%
\bibitem [{\citenamefont {Schlaepfer}\ \emph {et~al.}(2018)\citenamefont
  {Schlaepfer}, \citenamefont {Lucchini}, \citenamefont {Sato}, \citenamefont
  {Volkov}, \citenamefont {Kasmi}, \citenamefont {Hartmann}, \citenamefont
  {Rubio}, \citenamefont {Gallmann},\ and\ \citenamefont
  {Keller}}]{Schlaepfer2018natphys}%
  \BibitemOpen
  \bibfield  {author} {\bibinfo {author} {\bibfnamefont {F.}~\bibnamefont
  {Schlaepfer}}, \bibinfo {author} {\bibfnamefont {M.}~\bibnamefont
  {Lucchini}}, \bibinfo {author} {\bibfnamefont {S.~A.}\ \bibnamefont {Sato}},
  \bibinfo {author} {\bibfnamefont {M.}~\bibnamefont {Volkov}}, \bibinfo
  {author} {\bibfnamefont {L.}~\bibnamefont {Kasmi}}, \bibinfo {author}
  {\bibfnamefont {N.}~\bibnamefont {Hartmann}}, \bibinfo {author}
  {\bibfnamefont {A.}~\bibnamefont {Rubio}}, \bibinfo {author} {\bibfnamefont
  {L.}~\bibnamefont {Gallmann}}, \ and\ \bibinfo {author} {\bibfnamefont
  {U.}~\bibnamefont {Keller}},\ }\href@noop {} {\bibfield  {journal} {\bibinfo
  {journal} {Nature Physics (in press)}\ } (\bibinfo {year}
  {2018})}\BibitemShut {NoStop}%
\bibitem [{\citenamefont {Schultze}\ \emph {et~al.}(2010)\citenamefont
  {Schultze}, \citenamefont {Fie{\ss}}, \citenamefont {Karpowicz},
  \citenamefont {Gagnon}, \citenamefont {Korbman}, \citenamefont {Hofstetter},
  \citenamefont {Neppl}, \citenamefont {Cavalieri}, \citenamefont {Komninos},
  \citenamefont {Mercouris}, \citenamefont {Nicolaides}, \citenamefont
  {Pazourek}, \citenamefont {Nagele}, \citenamefont {Feist}, \citenamefont
  {Burgd{\"o}rfer}, \citenamefont {Azzeer}, \citenamefont {Ernstorfer},
  \citenamefont {Kienberger}, \citenamefont {Kleineberg}, \citenamefont
  {Goulielmakis}, \citenamefont {Krausz},\ and\ \citenamefont
  {Yakovlev}}]{Schultze1658}%
  \BibitemOpen
  \bibfield  {author} {\bibinfo {author} {\bibfnamefont {M.}~\bibnamefont
  {Schultze}}, \bibinfo {author} {\bibfnamefont {M.}~\bibnamefont {Fie{\ss}}},
  \bibinfo {author} {\bibfnamefont {N.}~\bibnamefont {Karpowicz}}, \bibinfo
  {author} {\bibfnamefont {J.}~\bibnamefont {Gagnon}}, \bibinfo {author}
  {\bibfnamefont {M.}~\bibnamefont {Korbman}}, \bibinfo {author} {\bibfnamefont
  {M.}~\bibnamefont {Hofstetter}}, \bibinfo {author} {\bibfnamefont
  {S.}~\bibnamefont {Neppl}}, \bibinfo {author} {\bibfnamefont {A.~L.}\
  \bibnamefont {Cavalieri}}, \bibinfo {author} {\bibfnamefont {Y.}~\bibnamefont
  {Komninos}}, \bibinfo {author} {\bibfnamefont {T.}~\bibnamefont {Mercouris}},
  \bibinfo {author} {\bibfnamefont {C.~A.}\ \bibnamefont {Nicolaides}},
  \bibinfo {author} {\bibfnamefont {R.}~\bibnamefont {Pazourek}}, \bibinfo
  {author} {\bibfnamefont {S.}~\bibnamefont {Nagele}}, \bibinfo {author}
  {\bibfnamefont {J.}~\bibnamefont {Feist}}, \bibinfo {author} {\bibfnamefont
  {J.}~\bibnamefont {Burgd{\"o}rfer}}, \bibinfo {author} {\bibfnamefont
  {A.~M.}\ \bibnamefont {Azzeer}}, \bibinfo {author} {\bibfnamefont
  {R.}~\bibnamefont {Ernstorfer}}, \bibinfo {author} {\bibfnamefont
  {R.}~\bibnamefont {Kienberger}}, \bibinfo {author} {\bibfnamefont
  {U.}~\bibnamefont {Kleineberg}}, \bibinfo {author} {\bibfnamefont
  {E.}~\bibnamefont {Goulielmakis}}, \bibinfo {author} {\bibfnamefont
  {F.}~\bibnamefont {Krausz}}, \ and\ \bibinfo {author} {\bibfnamefont {V.~S.}\
  \bibnamefont {Yakovlev}},\ }\href@noop {} {\bibfield  {journal} {\bibinfo
  {journal} {Science}\ }\textbf {\bibinfo {volume} {328}},\ \bibinfo {pages}
  {1658} (\bibinfo {year} {2010})}\BibitemShut {NoStop}%
\bibitem [{\citenamefont {Kl\"under}\ \emph {et~al.}(2011)\citenamefont
  {Kl\"under}, \citenamefont {Dahlstr\"om}, \citenamefont {Gisselbrecht},
  \citenamefont {Fordell}, \citenamefont {Swoboda}, \citenamefont {Gu\'enot},
  \citenamefont {Johnsson}, \citenamefont {Caillat}, \citenamefont
  {Mauritsson}, \citenamefont {Maquet}, \citenamefont {Ta\"{\i}eb},\ and\
  \citenamefont {L'Huillier}}]{PhysRevLett.106.143002}%
  \BibitemOpen
  \bibfield  {author} {\bibinfo {author} {\bibfnamefont {K.}~\bibnamefont
  {Kl\"under}}, \bibinfo {author} {\bibfnamefont {J.~M.}\ \bibnamefont
  {Dahlstr\"om}}, \bibinfo {author} {\bibfnamefont {M.}~\bibnamefont
  {Gisselbrecht}}, \bibinfo {author} {\bibfnamefont {T.}~\bibnamefont
  {Fordell}}, \bibinfo {author} {\bibfnamefont {M.}~\bibnamefont {Swoboda}},
  \bibinfo {author} {\bibfnamefont {D.}~\bibnamefont {Gu\'enot}}, \bibinfo
  {author} {\bibfnamefont {P.}~\bibnamefont {Johnsson}}, \bibinfo {author}
  {\bibfnamefont {J.}~\bibnamefont {Caillat}}, \bibinfo {author} {\bibfnamefont
  {J.}~\bibnamefont {Mauritsson}}, \bibinfo {author} {\bibfnamefont
  {A.}~\bibnamefont {Maquet}}, \bibinfo {author} {\bibfnamefont
  {R.}~\bibnamefont {Ta\"{\i}eb}}, \ and\ \bibinfo {author} {\bibfnamefont
  {A.}~\bibnamefont {L'Huillier}},\ }\href@noop {} {\bibfield  {journal}
  {\bibinfo  {journal} {Phys. Rev. Lett.}\ }\textbf {\bibinfo {volume} {106}},\
  \bibinfo {pages} {143002} (\bibinfo {year} {2011})}\BibitemShut {NoStop}%
\bibitem [{\citenamefont {Gu\'enot}\ \emph {et~al.}(2012)\citenamefont
  {Gu\'enot}, \citenamefont {Kl\"under}, \citenamefont {Arnold}, \citenamefont
  {Kroon}, \citenamefont {Dahlstr\"om}, \citenamefont {Miranda}, \citenamefont
  {Fordell}, \citenamefont {Gisselbrecht}, \citenamefont {Johnsson},
  \citenamefont {Mauritsson}, \citenamefont {Lindroth}, \citenamefont {Maquet},
  \citenamefont {Ta\"{\i}eb}, \citenamefont {L'Huillier},\ and\ \citenamefont
  {Kheifets}}]{PhysRevA.85.053424}%
  \BibitemOpen
  \bibfield  {author} {\bibinfo {author} {\bibfnamefont {D.}~\bibnamefont
  {Gu\'enot}}, \bibinfo {author} {\bibfnamefont {K.}~\bibnamefont {Kl\"under}},
  \bibinfo {author} {\bibfnamefont {C.~L.}\ \bibnamefont {Arnold}}, \bibinfo
  {author} {\bibfnamefont {D.}~\bibnamefont {Kroon}}, \bibinfo {author}
  {\bibfnamefont {J.~M.}\ \bibnamefont {Dahlstr\"om}}, \bibinfo {author}
  {\bibfnamefont {M.}~\bibnamefont {Miranda}}, \bibinfo {author} {\bibfnamefont
  {T.}~\bibnamefont {Fordell}}, \bibinfo {author} {\bibfnamefont
  {M.}~\bibnamefont {Gisselbrecht}}, \bibinfo {author} {\bibfnamefont
  {P.}~\bibnamefont {Johnsson}}, \bibinfo {author} {\bibfnamefont
  {J.}~\bibnamefont {Mauritsson}}, \bibinfo {author} {\bibfnamefont
  {E.}~\bibnamefont {Lindroth}}, \bibinfo {author} {\bibfnamefont
  {A.}~\bibnamefont {Maquet}}, \bibinfo {author} {\bibfnamefont
  {R.}~\bibnamefont {Ta\"{\i}eb}}, \bibinfo {author} {\bibfnamefont
  {A.}~\bibnamefont {L'Huillier}}, \ and\ \bibinfo {author} {\bibfnamefont
  {A.~S.}\ \bibnamefont {Kheifets}},\ }\href@noop {} {\bibfield  {journal}
  {\bibinfo  {journal} {Phys. Rev. A}\ }\textbf {\bibinfo {volume} {85}},\
  \bibinfo {pages} {053424} (\bibinfo {year} {2012})}\BibitemShut {NoStop}%
\bibitem [{\citenamefont {Isinger}\ \emph {et~al.}(2017)\citenamefont
  {Isinger}, \citenamefont {Squibb}, \citenamefont {Busto}, \citenamefont
  {Zhong}, \citenamefont {Harth}, \citenamefont {Kroon}, \citenamefont {Nandi},
  \citenamefont {Arnold}, \citenamefont {Miranda}, \citenamefont
  {Dahlstr{\"o}m}, \citenamefont {Lindroth}, \citenamefont {Feifel},
  \citenamefont {Gisselbrecht},\ and\ \citenamefont
  {L{\textquoteright}Huillier}}]{Isinger893}%
  \BibitemOpen
  \bibfield  {author} {\bibinfo {author} {\bibfnamefont {M.}~\bibnamefont
  {Isinger}}, \bibinfo {author} {\bibfnamefont {R.~J.}\ \bibnamefont {Squibb}},
  \bibinfo {author} {\bibfnamefont {D.}~\bibnamefont {Busto}}, \bibinfo
  {author} {\bibfnamefont {S.}~\bibnamefont {Zhong}}, \bibinfo {author}
  {\bibfnamefont {A.}~\bibnamefont {Harth}}, \bibinfo {author} {\bibfnamefont
  {D.}~\bibnamefont {Kroon}}, \bibinfo {author} {\bibfnamefont
  {S.}~\bibnamefont {Nandi}}, \bibinfo {author} {\bibfnamefont {C.~L.}\
  \bibnamefont {Arnold}}, \bibinfo {author} {\bibfnamefont {M.}~\bibnamefont
  {Miranda}}, \bibinfo {author} {\bibfnamefont {J.~M.}\ \bibnamefont
  {Dahlstr{\"o}m}}, \bibinfo {author} {\bibfnamefont {E.}~\bibnamefont
  {Lindroth}}, \bibinfo {author} {\bibfnamefont {R.}~\bibnamefont {Feifel}},
  \bibinfo {author} {\bibfnamefont {M.}~\bibnamefont {Gisselbrecht}}, \ and\
  \bibinfo {author} {\bibfnamefont {A.}~\bibnamefont
  {L{\textquoteright}Huillier}},\ }\href@noop {} {\bibfield  {journal}
  {\bibinfo  {journal} {Science}\ }\textbf {\bibinfo {volume} {358}},\ \bibinfo
  {pages} {893} (\bibinfo {year} {2017})}\BibitemShut {NoStop}%
\bibitem [{\citenamefont {Cavalieri}\ \emph {et~al.}(2007)\citenamefont
  {Cavalieri}, \citenamefont {M{\"u}ller}, \citenamefont {Uphues},
  \citenamefont {Yakovlev}, \citenamefont {Baltu{\v{s}}ka}, \citenamefont
  {Horvath}, \citenamefont {Schmidt}, \citenamefont {Bl{\"u}mel}, \citenamefont
  {Holzwarth}, \citenamefont {Hendel} \emph
  {et~al.}}]{cavalieri2007attosecond}%
  \BibitemOpen
  \bibfield  {author} {\bibinfo {author} {\bibfnamefont {A.~L.}\ \bibnamefont
  {Cavalieri}}, \bibinfo {author} {\bibfnamefont {N.}~\bibnamefont
  {M{\"u}ller}}, \bibinfo {author} {\bibfnamefont {T.}~\bibnamefont {Uphues}},
  \bibinfo {author} {\bibfnamefont {V.~S.}\ \bibnamefont {Yakovlev}}, \bibinfo
  {author} {\bibfnamefont {A.}~\bibnamefont {Baltu{\v{s}}ka}}, \bibinfo
  {author} {\bibfnamefont {B.}~\bibnamefont {Horvath}}, \bibinfo {author}
  {\bibfnamefont {B.}~\bibnamefont {Schmidt}}, \bibinfo {author} {\bibfnamefont
  {L.}~\bibnamefont {Bl{\"u}mel}}, \bibinfo {author} {\bibfnamefont
  {R.}~\bibnamefont {Holzwarth}}, \bibinfo {author} {\bibfnamefont
  {S.}~\bibnamefont {Hendel}},  \emph {et~al.},\ }\href@noop {} {\bibfield
  {journal} {\bibinfo  {journal} {Nature}\ }\textbf {\bibinfo {volume} {449}},\
  \bibinfo {pages} {1029} (\bibinfo {year} {2007})}\BibitemShut {NoStop}%
\bibitem [{\citenamefont {Locher}\ \emph {et~al.}(2015)\citenamefont {Locher},
  \citenamefont {Castiglioni}, \citenamefont {Lucchini}, \citenamefont {Greif},
  \citenamefont {Gallmann}, \citenamefont {Osterwalder}, \citenamefont
  {Hengsberger},\ and\ \citenamefont {Keller}}]{Locher:15}%
  \BibitemOpen
  \bibfield  {author} {\bibinfo {author} {\bibfnamefont {R.}~\bibnamefont
  {Locher}}, \bibinfo {author} {\bibfnamefont {L.}~\bibnamefont {Castiglioni}},
  \bibinfo {author} {\bibfnamefont {M.}~\bibnamefont {Lucchini}}, \bibinfo
  {author} {\bibfnamefont {M.}~\bibnamefont {Greif}}, \bibinfo {author}
  {\bibfnamefont {L.}~\bibnamefont {Gallmann}}, \bibinfo {author}
  {\bibfnamefont {J.}~\bibnamefont {Osterwalder}}, \bibinfo {author}
  {\bibfnamefont {M.}~\bibnamefont {Hengsberger}}, \ and\ \bibinfo {author}
  {\bibfnamefont {U.}~\bibnamefont {Keller}},\ }\href@noop {} {\bibfield
  {journal} {\bibinfo  {journal} {Optica}\ }\textbf {\bibinfo {volume} {2}},\
  \bibinfo {pages} {405} (\bibinfo {year} {2015})}\BibitemShut {NoStop}%
\bibitem [{\citenamefont {Neppl}\ \emph {et~al.}(2015)\citenamefont {Neppl},
  \citenamefont {Ernstorfer}, \citenamefont {Cavalieri}, \citenamefont
  {Lemell}, \citenamefont {Wachter}, \citenamefont {Magerl}, \citenamefont
  {Bothschafter}, \citenamefont {Jobst}, \citenamefont {Hofstetter},
  \citenamefont {Kleineberg}, \citenamefont {Barth}, \citenamefont {Menzel},
  \citenamefont {Burgd\"orfer}, \citenamefont {Feulner}, \citenamefont
  {Krausz},\ and\ \citenamefont {Kienberger}}]{neppl2015direct}%
  \BibitemOpen
  \bibfield  {author} {\bibinfo {author} {\bibfnamefont {S.}~\bibnamefont
  {Neppl}}, \bibinfo {author} {\bibfnamefont {R.}~\bibnamefont {Ernstorfer}},
  \bibinfo {author} {\bibfnamefont {A.~L.}\ \bibnamefont {Cavalieri}}, \bibinfo
  {author} {\bibfnamefont {C.}~\bibnamefont {Lemell}}, \bibinfo {author}
  {\bibfnamefont {G.}~\bibnamefont {Wachter}}, \bibinfo {author} {\bibfnamefont
  {E.}~\bibnamefont {Magerl}}, \bibinfo {author} {\bibfnamefont {E.~M.}\
  \bibnamefont {Bothschafter}}, \bibinfo {author} {\bibfnamefont
  {M.}~\bibnamefont {Jobst}}, \bibinfo {author} {\bibfnamefont
  {M.}~\bibnamefont {Hofstetter}}, \bibinfo {author} {\bibfnamefont
  {U.}~\bibnamefont {Kleineberg}}, \bibinfo {author} {\bibfnamefont {J.~V.}\
  \bibnamefont {Barth}}, \bibinfo {author} {\bibfnamefont {D.}~\bibnamefont
  {Menzel}}, \bibinfo {author} {\bibfnamefont {J.}~\bibnamefont
  {Burgd\"orfer}}, \bibinfo {author} {\bibfnamefont {P.}~\bibnamefont
  {Feulner}}, \bibinfo {author} {\bibfnamefont {F.}~\bibnamefont {Krausz}}, \
  and\ \bibinfo {author} {\bibfnamefont {R.}~\bibnamefont {Kienberger}},\
  }\href@noop {} {\bibfield  {journal} {\bibinfo  {journal} {Nature}\ }\textbf
  {\bibinfo {volume} {517}},\ \bibinfo {pages} {342} (\bibinfo {year}
  {2015})}\BibitemShut {NoStop}%
\bibitem [{\citenamefont {Moore}\ \emph {et~al.}(2011)\citenamefont {Moore},
  \citenamefont {Lysaght}, \citenamefont {Parker}, \citenamefont {van~der
  Hart},\ and\ \citenamefont {Taylor}}]{PhysRevA.84.061404}%
  \BibitemOpen
  \bibfield  {author} {\bibinfo {author} {\bibfnamefont {L.~R.}\ \bibnamefont
  {Moore}}, \bibinfo {author} {\bibfnamefont {M.~A.}\ \bibnamefont {Lysaght}},
  \bibinfo {author} {\bibfnamefont {J.~S.}\ \bibnamefont {Parker}}, \bibinfo
  {author} {\bibfnamefont {H.~W.}\ \bibnamefont {van~der Hart}}, \ and\
  \bibinfo {author} {\bibfnamefont {K.~T.}\ \bibnamefont {Taylor}},\
  }\href@noop {} {\bibfield  {journal} {\bibinfo  {journal} {Phys. Rev. A}\
  }\textbf {\bibinfo {volume} {84}},\ \bibinfo {pages} {061404} (\bibinfo
  {year} {2011})}\BibitemShut {NoStop}%
\bibitem [{\citenamefont {Dahlstr\"om}\ \emph {et~al.}(2012)\citenamefont
  {Dahlstr\"om}, \citenamefont {Carette},\ and\ \citenamefont
  {Lindroth}}]{PhysRevA.86.061402}%
  \BibitemOpen
  \bibfield  {author} {\bibinfo {author} {\bibfnamefont {J.~M.}\ \bibnamefont
  {Dahlstr\"om}}, \bibinfo {author} {\bibfnamefont {T.}~\bibnamefont
  {Carette}}, \ and\ \bibinfo {author} {\bibfnamefont {E.}~\bibnamefont
  {Lindroth}},\ }\href@noop {} {\bibfield  {journal} {\bibinfo  {journal}
  {Phys. Rev. A}\ }\textbf {\bibinfo {volume} {86}},\ \bibinfo {pages} {061402}
  (\bibinfo {year} {2012})}\BibitemShut {NoStop}%
\bibitem [{\citenamefont {Feist}\ \emph {et~al.}(2014)\citenamefont {Feist},
  \citenamefont {Zatsarinny}, \citenamefont {Nagele}, \citenamefont {Pazourek},
  \citenamefont {Burgd\"orfer}, \citenamefont {Guan}, \citenamefont
  {Bartschat},\ and\ \citenamefont {Schneider}}]{PhysRevA.89.033417}%
  \BibitemOpen
  \bibfield  {author} {\bibinfo {author} {\bibfnamefont {J.}~\bibnamefont
  {Feist}}, \bibinfo {author} {\bibfnamefont {O.}~\bibnamefont {Zatsarinny}},
  \bibinfo {author} {\bibfnamefont {S.}~\bibnamefont {Nagele}}, \bibinfo
  {author} {\bibfnamefont {R.}~\bibnamefont {Pazourek}}, \bibinfo {author}
  {\bibfnamefont {J.}~\bibnamefont {Burgd\"orfer}}, \bibinfo {author}
  {\bibfnamefont {X.}~\bibnamefont {Guan}}, \bibinfo {author} {\bibfnamefont
  {K.}~\bibnamefont {Bartschat}}, \ and\ \bibinfo {author} {\bibfnamefont
  {B.~I.}\ \bibnamefont {Schneider}},\ }\href@noop {} {\bibfield  {journal}
  {\bibinfo  {journal} {Phys. Rev. A}\ }\textbf {\bibinfo {volume} {89}},\
  \bibinfo {pages} {033417} (\bibinfo {year} {2014})}\BibitemShut {NoStop}%
\bibitem [{\citenamefont {Otobe}(2012)}]{doi:10.1063/1.4716192}%
  \BibitemOpen
  \bibfield  {author} {\bibinfo {author} {\bibfnamefont {T.}~\bibnamefont
  {Otobe}},\ }\href@noop {} {\bibfield  {journal} {\bibinfo  {journal} {Journal
  of Applied Physics}\ }\textbf {\bibinfo {volume} {111}},\ \bibinfo {pages}
  {093112} (\bibinfo {year} {2012})}\BibitemShut {NoStop}%
\bibitem [{\citenamefont {Tancogne-Dejean}\ \emph {et~al.}(2017)\citenamefont
  {Tancogne-Dejean}, \citenamefont {M\"ucke}, \citenamefont {K\"artner},\ and\
  \citenamefont {Rubio}}]{PhysRevLett.118.087403}%
  \BibitemOpen
  \bibfield  {author} {\bibinfo {author} {\bibfnamefont {N.}~\bibnamefont
  {Tancogne-Dejean}}, \bibinfo {author} {\bibfnamefont {O.~D.}\ \bibnamefont
  {M\"ucke}}, \bibinfo {author} {\bibfnamefont {F.~X.}\ \bibnamefont
  {K\"artner}}, \ and\ \bibinfo {author} {\bibfnamefont {A.}~\bibnamefont
  {Rubio}},\ }\href@noop {} {\bibfield  {journal} {\bibinfo  {journal} {Phys.
  Rev. Lett.}\ }\textbf {\bibinfo {volume} {118}},\ \bibinfo {pages} {087403}
  (\bibinfo {year} {2017})}\BibitemShut {NoStop}%
\bibitem [{\citenamefont {Floss}\ \emph {et~al.}(2018)\citenamefont {Floss},
  \citenamefont {Lemell}, \citenamefont {Wachter}, \citenamefont {Smejkal},
  \citenamefont {Sato}, \citenamefont {Tong}, \citenamefont {Yabana},\ and\
  \citenamefont {Burgd\"orfer}}]{PhysRevA.97.011401}%
  \BibitemOpen
  \bibfield  {author} {\bibinfo {author} {\bibfnamefont {I.}~\bibnamefont
  {Floss}}, \bibinfo {author} {\bibfnamefont {C.}~\bibnamefont {Lemell}},
  \bibinfo {author} {\bibfnamefont {G.}~\bibnamefont {Wachter}}, \bibinfo
  {author} {\bibfnamefont {V.}~\bibnamefont {Smejkal}}, \bibinfo {author}
  {\bibfnamefont {S.~A.}\ \bibnamefont {Sato}}, \bibinfo {author}
  {\bibfnamefont {X.-M.}\ \bibnamefont {Tong}}, \bibinfo {author}
  {\bibfnamefont {K.}~\bibnamefont {Yabana}}, \ and\ \bibinfo {author}
  {\bibfnamefont {J.}~\bibnamefont {Burgd\"orfer}},\ }\href@noop {} {\bibfield
  {journal} {\bibinfo  {journal} {Phys. Rev. A}\ }\textbf {\bibinfo {volume}
  {97}},\ \bibinfo {pages} {011401} (\bibinfo {year} {2018})}\BibitemShut
  {NoStop}%
\bibitem [{\citenamefont {Sato}\ \emph {et~al.}(2015)\citenamefont {Sato},
  \citenamefont {Yabana}, \citenamefont {Shinohara}, \citenamefont {Otobe},
  \citenamefont {Lee},\ and\ \citenamefont {Bertsch}}]{PhysRevB.92.205413}%
  \BibitemOpen
  \bibfield  {author} {\bibinfo {author} {\bibfnamefont {S.~A.}\ \bibnamefont
  {Sato}}, \bibinfo {author} {\bibfnamefont {K.}~\bibnamefont {Yabana}},
  \bibinfo {author} {\bibfnamefont {Y.}~\bibnamefont {Shinohara}}, \bibinfo
  {author} {\bibfnamefont {T.}~\bibnamefont {Otobe}}, \bibinfo {author}
  {\bibfnamefont {K.-M.}\ \bibnamefont {Lee}}, \ and\ \bibinfo {author}
  {\bibfnamefont {G.~F.}\ \bibnamefont {Bertsch}},\ }\href@noop {} {\bibfield
  {journal} {\bibinfo  {journal} {Phys. Rev. B}\ }\textbf {\bibinfo {volume}
  {92}},\ \bibinfo {pages} {205413} (\bibinfo {year} {2015})}\BibitemShut
  {NoStop}%
\bibitem [{\citenamefont {Shin}\ \emph {et~al.}(2018)\citenamefont {Shin},
  \citenamefont {H\"ubener}, \citenamefont {De~Giovannini}, \citenamefont
  {Jin}, \citenamefont {Rubio},\ and\ \citenamefont {Park}}]{noejung2018}%
  \BibitemOpen
  \bibfield  {author} {\bibinfo {author} {\bibfnamefont {D.}~\bibnamefont
  {Shin}}, \bibinfo {author} {\bibfnamefont {H.}~\bibnamefont {H\"ubener}},
  \bibinfo {author} {\bibfnamefont {U.}~\bibnamefont {De~Giovannini}}, \bibinfo
  {author} {\bibfnamefont {H.}~\bibnamefont {Jin}}, \bibinfo {author}
  {\bibfnamefont {A.}~\bibnamefont {Rubio}}, \ and\ \bibinfo {author}
  {\bibfnamefont {N.}~\bibnamefont {Park}},\ }\href@noop {} {\bibfield
  {journal} {\bibinfo  {journal} {Nature communications}\ }\textbf {\bibinfo
  {volume} {9}},\ \bibinfo {pages} {638} (\bibinfo {year} {2018})}\BibitemShut
  {NoStop}%
\bibitem [{\citenamefont {De~Giovannini}\ \emph {et~al.}(2016)\citenamefont
  {De~Giovannini}, \citenamefont {H{\"u}bener},\ and\ \citenamefont
  {Rubio}}]{deGiovannini:2016bb}%
  \BibitemOpen
  \bibfield  {author} {\bibinfo {author} {\bibfnamefont {U.}~\bibnamefont
  {De~Giovannini}}, \bibinfo {author} {\bibfnamefont {H.}~\bibnamefont
  {H{\"u}bener}}, \ and\ \bibinfo {author} {\bibfnamefont {A.}~\bibnamefont
  {Rubio}},\ }\href@noop {} {\bibfield  {journal} {\bibinfo  {journal} {Journal
  of Chemical Theory and Computation}\ }\textbf {\bibinfo {volume} {13}},\
  \bibinfo {pages} {265} (\bibinfo {year} {2016})}\BibitemShut {NoStop}%
\bibitem [{\citenamefont {Pazourek}\ \emph {et~al.}(2015)\citenamefont
  {Pazourek}, \citenamefont {Nagele},\ and\ \citenamefont
  {Burgd\"orfer}}]{RevModPhys.87.765}%
  \BibitemOpen
  \bibfield  {author} {\bibinfo {author} {\bibfnamefont {R.}~\bibnamefont
  {Pazourek}}, \bibinfo {author} {\bibfnamefont {S.}~\bibnamefont {Nagele}}, \
  and\ \bibinfo {author} {\bibfnamefont {J.}~\bibnamefont {Burgd\"orfer}},\
  }\href@noop {} {\bibfield  {journal} {\bibinfo  {journal} {Rev. Mod. Phys.}\
  }\textbf {\bibinfo {volume} {87}},\ \bibinfo {pages} {765} (\bibinfo {year}
  {2015})}\BibitemShut {NoStop}%
\bibitem [{\citenamefont {Cattaneo}\ \emph {et~al.}(2016)\citenamefont
  {Cattaneo}, \citenamefont {Vos}, \citenamefont {Lucchini}, \citenamefont
  {Gallmann}, \citenamefont {Cirelli},\ and\ \citenamefont
  {Keller}}]{Cattaneo:16}%
  \BibitemOpen
  \bibfield  {author} {\bibinfo {author} {\bibfnamefont {L.}~\bibnamefont
  {Cattaneo}}, \bibinfo {author} {\bibfnamefont {J.}~\bibnamefont {Vos}},
  \bibinfo {author} {\bibfnamefont {M.}~\bibnamefont {Lucchini}}, \bibinfo
  {author} {\bibfnamefont {L.}~\bibnamefont {Gallmann}}, \bibinfo {author}
  {\bibfnamefont {C.}~\bibnamefont {Cirelli}}, \ and\ \bibinfo {author}
  {\bibfnamefont {U.}~\bibnamefont {Keller}},\ }\href@noop {} {\bibfield
  {journal} {\bibinfo  {journal} {Opt. Express}\ }\textbf {\bibinfo {volume}
  {24}},\ \bibinfo {pages} {29060} (\bibinfo {year} {2016})}\BibitemShut
  {NoStop}%
\bibitem [{\citenamefont {Kasmi}\ \emph {et~al.}(2017)\citenamefont {Kasmi},
  \citenamefont {Lucchini}, \citenamefont {Castiglioni}, \citenamefont
  {Kliuiev}, \citenamefont {Osterwalder}, \citenamefont {Hengsberger},
  \citenamefont {Gallmann}, \citenamefont {Kr\"{u}ger},\ and\ \citenamefont
  {Keller}}]{Kasmi:17}%
  \BibitemOpen
  \bibfield  {author} {\bibinfo {author} {\bibfnamefont {L.}~\bibnamefont
  {Kasmi}}, \bibinfo {author} {\bibfnamefont {M.}~\bibnamefont {Lucchini}},
  \bibinfo {author} {\bibfnamefont {L.}~\bibnamefont {Castiglioni}}, \bibinfo
  {author} {\bibfnamefont {P.}~\bibnamefont {Kliuiev}}, \bibinfo {author}
  {\bibfnamefont {J.}~\bibnamefont {Osterwalder}}, \bibinfo {author}
  {\bibfnamefont {M.}~\bibnamefont {Hengsberger}}, \bibinfo {author}
  {\bibfnamefont {L.}~\bibnamefont {Gallmann}}, \bibinfo {author}
  {\bibfnamefont {P.}~\bibnamefont {Kr\"{u}ger}}, \ and\ \bibinfo {author}
  {\bibfnamefont {U.}~\bibnamefont {Keller}},\ }\href@noop {} {\bibfield
  {journal} {\bibinfo  {journal} {Optica}\ }\textbf {\bibinfo {volume} {4}},\
  \bibinfo {pages} {1492} (\bibinfo {year} {2017})}\BibitemShut {NoStop}%
\bibitem [{\citenamefont {Ceperley}\ and\ \citenamefont
  {Alder}(1980)}]{PhysRevLett.45.566}%
  \BibitemOpen
  \bibfield  {author} {\bibinfo {author} {\bibfnamefont {D.~M.}\ \bibnamefont
  {Ceperley}}\ and\ \bibinfo {author} {\bibfnamefont {B.~J.}\ \bibnamefont
  {Alder}},\ }\href@noop {} {\bibfield  {journal} {\bibinfo  {journal} {Phys.
  Rev. Lett.}\ }\textbf {\bibinfo {volume} {45}},\ \bibinfo {pages} {566}
  (\bibinfo {year} {1980})}\BibitemShut {NoStop}%
\bibitem [{\citenamefont {Perdew}\ and\ \citenamefont
  {Zunger}(1981)}]{PhysRevB.23.5048}%
  \BibitemOpen
  \bibfield  {author} {\bibinfo {author} {\bibfnamefont {J.~P.}\ \bibnamefont
  {Perdew}}\ and\ \bibinfo {author} {\bibfnamefont {A.}~\bibnamefont
  {Zunger}},\ }\href@noop {} {\bibfield  {journal} {\bibinfo  {journal} {Phys.
  Rev. B}\ }\textbf {\bibinfo {volume} {23}},\ \bibinfo {pages} {5048}
  (\bibinfo {year} {1981})}\BibitemShut {NoStop}%
\bibitem [{\citenamefont {Legrand}\ \emph {et~al.}(2002)\citenamefont
  {Legrand}, \citenamefont {Suraud},\ and\ \citenamefont
  {Reinhard}}]{Legrand:2002jf}%
  \BibitemOpen
  \bibfield  {author} {\bibinfo {author} {\bibfnamefont {C.}~\bibnamefont
  {Legrand}}, \bibinfo {author} {\bibfnamefont {E.}~\bibnamefont {Suraud}}, \
  and\ \bibinfo {author} {\bibfnamefont {P.~G.}\ \bibnamefont {Reinhard}},\
  }\href@noop {} {\bibfield  {journal} {\bibinfo  {journal} {Journal of Physics
  B: Atomic, Molecular and Optical Physics}\ }\textbf {\bibinfo {volume}
  {35}},\ \bibinfo {pages} {1115} (\bibinfo {year} {2002})}\BibitemShut
  {NoStop}%
\bibitem [{\citenamefont {Tao}\ and\ \citenamefont
  {Scrinzi}(2012)}]{Tao:2012ev}%
  \BibitemOpen
  \bibfield  {author} {\bibinfo {author} {\bibfnamefont {L.}~\bibnamefont
  {Tao}}\ and\ \bibinfo {author} {\bibfnamefont {A.}~\bibnamefont {Scrinzi}},\
  }\href@noop {} {\bibfield  {journal} {\bibinfo  {journal} {New Journal of
  Physics}\ }\textbf {\bibinfo {volume} {14}},\ \bibinfo {pages} {013021}
  (\bibinfo {year} {2012})}\BibitemShut {NoStop}%
\bibitem [{\citenamefont {Wopperer}\ \emph {et~al.}(2017)\citenamefont
  {Wopperer}, \citenamefont {De~Giovannini},\ and\ \citenamefont
  {Rubio}}]{Wopperer:2017bm}%
  \BibitemOpen
  \bibfield  {author} {\bibinfo {author} {\bibfnamefont {P.}~\bibnamefont
  {Wopperer}}, \bibinfo {author} {\bibfnamefont {U.}~\bibnamefont
  {De~Giovannini}}, \ and\ \bibinfo {author} {\bibfnamefont {A.}~\bibnamefont
  {Rubio}},\ }\href@noop {} {\bibfield  {journal} {\bibinfo  {journal} {The
  European Physical Journal B}\ }\textbf {\bibinfo {volume} {90}},\ \bibinfo
  {pages} {1307} (\bibinfo {year} {2017})}\BibitemShut {NoStop}%
\bibitem [{\citenamefont {De~Giovannini}\ \emph {et~al.}(2015)\citenamefont
  {De~Giovannini}, \citenamefont {Larsen}, \citenamefont {Rubio},\ and\
  \citenamefont {Rubio}}]{DeGiovannini:2015jt}%
  \BibitemOpen
  \bibfield  {author} {\bibinfo {author} {\bibfnamefont {U.}~\bibnamefont
  {De~Giovannini}}, \bibinfo {author} {\bibfnamefont {A.~H.}\ \bibnamefont
  {Larsen}}, \bibinfo {author} {\bibfnamefont {A.}~\bibnamefont {Rubio}}, \
  and\ \bibinfo {author} {\bibfnamefont {A.}~\bibnamefont {Rubio}},\
  }\href@noop {} {\bibfield  {journal} {\bibinfo  {journal} {The European
  Physical Journal B}\ }\textbf {\bibinfo {volume} {88}},\ \bibinfo {pages} {1}
  (\bibinfo {year} {2015})}\BibitemShut {NoStop}%
\bibitem [{\citenamefont {Hartwigsen}\ \emph {et~al.}(1998)\citenamefont
  {Hartwigsen}, \citenamefont {Goedecker},\ and\ \citenamefont
  {Hutter}}]{PhysRevB.58.3641}%
  \BibitemOpen
  \bibfield  {author} {\bibinfo {author} {\bibfnamefont {C.}~\bibnamefont
  {Hartwigsen}}, \bibinfo {author} {\bibfnamefont {S.}~\bibnamefont
  {Goedecker}}, \ and\ \bibinfo {author} {\bibfnamefont {J.}~\bibnamefont
  {Hutter}},\ }\href@noop {} {\bibfield  {journal} {\bibinfo  {journal} {Phys.
  Rev. B}\ }\textbf {\bibinfo {volume} {58}},\ \bibinfo {pages} {3641}
  (\bibinfo {year} {1998})}\BibitemShut {NoStop}%
\bibitem [{\citenamefont {Andrade}\ \emph {et~al.}(2015)\citenamefont
  {Andrade}, \citenamefont {Strubbe}, \citenamefont {De~Giovannini},
  \citenamefont {Larsen}, \citenamefont {Oliveira}, \citenamefont
  {Alberdi-Rodriguez}, \citenamefont {Varas}, \citenamefont {Theophilou},
  \citenamefont {Helbig}, \citenamefont {Verstraete}, \citenamefont {Stella},
  \citenamefont {Nogueira}, \citenamefont {Aspuru-Guzik}, \citenamefont
  {Castro}, \citenamefont {Marques}, \citenamefont {Rubio},\ and\ \citenamefont
  {Rubio}}]{Strubbe:2015iz}%
  \BibitemOpen
  \bibfield  {author} {\bibinfo {author} {\bibfnamefont {X.}~\bibnamefont
  {Andrade}}, \bibinfo {author} {\bibfnamefont {D.}~\bibnamefont {Strubbe}},
  \bibinfo {author} {\bibfnamefont {U.}~\bibnamefont {De~Giovannini}}, \bibinfo
  {author} {\bibfnamefont {A.~H.}\ \bibnamefont {Larsen}}, \bibinfo {author}
  {\bibfnamefont {M.~J.~T.}\ \bibnamefont {Oliveira}}, \bibinfo {author}
  {\bibfnamefont {J.}~\bibnamefont {Alberdi-Rodriguez}}, \bibinfo {author}
  {\bibfnamefont {A.}~\bibnamefont {Varas}}, \bibinfo {author} {\bibfnamefont
  {I.}~\bibnamefont {Theophilou}}, \bibinfo {author} {\bibfnamefont
  {N.}~\bibnamefont {Helbig}}, \bibinfo {author} {\bibfnamefont {M.~J.}\
  \bibnamefont {Verstraete}}, \bibinfo {author} {\bibfnamefont
  {L.}~\bibnamefont {Stella}}, \bibinfo {author} {\bibfnamefont
  {F.}~\bibnamefont {Nogueira}}, \bibinfo {author} {\bibfnamefont
  {A.}~\bibnamefont {Aspuru-Guzik}}, \bibinfo {author} {\bibfnamefont
  {A.}~\bibnamefont {Castro}}, \bibinfo {author} {\bibfnamefont {M.~A.~L.}\
  \bibnamefont {Marques}}, \bibinfo {author} {\bibfnamefont {A.}~\bibnamefont
  {Rubio}}, \ and\ \bibinfo {author} {\bibfnamefont {A.}~\bibnamefont
  {Rubio}},\ }\href@noop {} {\bibfield  {journal} {\bibinfo  {journal}
  {Physical Chemistry Chemical Physics}\ }\textbf {\bibinfo {volume} {17}},\
  \bibinfo {pages} {31371} (\bibinfo {year} {2015})}\BibitemShut {NoStop}%
\bibitem [{\citenamefont {Magrakvelidze}\ \emph {et~al.}(2015)\citenamefont
  {Magrakvelidze}, \citenamefont {Madjet}, \citenamefont {Dixit}, \citenamefont
  {Ivanov},\ and\ \citenamefont {Chakraborty}}]{PhysRevA.91.063415}%
  \BibitemOpen
  \bibfield  {author} {\bibinfo {author} {\bibfnamefont {M.}~\bibnamefont
  {Magrakvelidze}}, \bibinfo {author} {\bibfnamefont {M.~E.-A.}\ \bibnamefont
  {Madjet}}, \bibinfo {author} {\bibfnamefont {G.}~\bibnamefont {Dixit}},
  \bibinfo {author} {\bibfnamefont {M.}~\bibnamefont {Ivanov}}, \ and\ \bibinfo
  {author} {\bibfnamefont {H.~S.}\ \bibnamefont {Chakraborty}},\ }\href@noop {}
  {\bibfield  {journal} {\bibinfo  {journal} {Phys. Rev. A}\ }\textbf {\bibinfo
  {volume} {91}},\ \bibinfo {pages} {063415} (\bibinfo {year}
  {2015})}\BibitemShut {NoStop}%
\bibitem [{\citenamefont {Wigner}(1955)}]{PhysRev.98.145}%
  \BibitemOpen
  \bibfield  {author} {\bibinfo {author} {\bibfnamefont {E.~P.}\ \bibnamefont
  {Wigner}},\ }\href@noop {} {\bibfield  {journal} {\bibinfo  {journal} {Phys.
  Rev.}\ }\textbf {\bibinfo {volume} {98}},\ \bibinfo {pages} {145} (\bibinfo
  {year} {1955})}\BibitemShut {NoStop}%
\bibitem [{\citenamefont {Smith}(1960)}]{PhysRev.118.349}%
  \BibitemOpen
  \bibfield  {author} {\bibinfo {author} {\bibfnamefont {F.~T.}\ \bibnamefont
  {Smith}},\ }\href@noop {} {\bibfield  {journal} {\bibinfo  {journal} {Phys.
  Rev.}\ }\textbf {\bibinfo {volume} {118}},\ \bibinfo {pages} {349} (\bibinfo
  {year} {1960})}\BibitemShut {NoStop}%
\bibitem [{\citenamefont {Dahlström}\ \emph {et~al.}(2013)\citenamefont
  {Dahlström}, \citenamefont {Guénot}, \citenamefont {Klünder},
  \citenamefont {Gisselbrecht}, \citenamefont {Mauritsson}, \citenamefont
  {L’Huillier}, \citenamefont {Maquet},\ and\ \citenamefont
  {Taïeb}}]{DAHLSTROM201353}%
  \BibitemOpen
  \bibfield  {author} {\bibinfo {author} {\bibfnamefont {J.}~\bibnamefont
  {Dahlström}}, \bibinfo {author} {\bibfnamefont {D.}~\bibnamefont {Guénot}},
  \bibinfo {author} {\bibfnamefont {K.}~\bibnamefont {Klünder}}, \bibinfo
  {author} {\bibfnamefont {M.}~\bibnamefont {Gisselbrecht}}, \bibinfo {author}
  {\bibfnamefont {J.}~\bibnamefont {Mauritsson}}, \bibinfo {author}
  {\bibfnamefont {A.}~\bibnamefont {L’Huillier}}, \bibinfo {author}
  {\bibfnamefont {A.}~\bibnamefont {Maquet}}, \ and\ \bibinfo {author}
  {\bibfnamefont {R.}~\bibnamefont {Taïeb}},\ }\href@noop {} {\bibfield
  {journal} {\bibinfo  {journal} {Chemical Physics}\ }\textbf {\bibinfo
  {volume} {414}},\ \bibinfo {pages} {53 } (\bibinfo {year}
  {2013})}\BibitemShut {NoStop}%
\bibitem [{\citenamefont {M\"obus}\ \emph {et~al.}(1993)\citenamefont
  {M\"obus}, \citenamefont {Magel}, \citenamefont {Schartner}, \citenamefont
  {Langer}, \citenamefont {Becker}, \citenamefont {Wildberger},\ and\
  \citenamefont {Schmoranzer}}]{PhysRevA.47.3888}%
  \BibitemOpen
  \bibfield  {author} {\bibinfo {author} {\bibfnamefont {B.}~\bibnamefont
  {M\"obus}}, \bibinfo {author} {\bibfnamefont {B.}~\bibnamefont {Magel}},
  \bibinfo {author} {\bibfnamefont {K.-H.}\ \bibnamefont {Schartner}}, \bibinfo
  {author} {\bibfnamefont {B.}~\bibnamefont {Langer}}, \bibinfo {author}
  {\bibfnamefont {U.}~\bibnamefont {Becker}}, \bibinfo {author} {\bibfnamefont
  {M.}~\bibnamefont {Wildberger}}, \ and\ \bibinfo {author} {\bibfnamefont
  {H.}~\bibnamefont {Schmoranzer}},\ }\href@noop {} {\bibfield  {journal}
  {\bibinfo  {journal} {Phys. Rev. A}\ }\textbf {\bibinfo {volume} {47}},\
  \bibinfo {pages} {3888} (\bibinfo {year} {1993})}\BibitemShut {NoStop}%
\bibitem [{\citenamefont {Crawford-Uranga}\ \emph {et~al.}(2014)\citenamefont
  {Crawford-Uranga}, \citenamefont {De~Giovannini}, \citenamefont
  {R\"as\"anen}, \citenamefont {Oliveira}, \citenamefont {Mowbray},
  \citenamefont {Nikolopoulos}, \citenamefont {Karamatskos}, \citenamefont
  {Markellos}, \citenamefont {Lambropoulos}, \citenamefont {Kurth},\ and\
  \citenamefont {Rubio}}]{PhysRevA.90.033412}%
  \BibitemOpen
  \bibfield  {author} {\bibinfo {author} {\bibfnamefont {A.}~\bibnamefont
  {Crawford-Uranga}}, \bibinfo {author} {\bibfnamefont {U.}~\bibnamefont
  {De~Giovannini}}, \bibinfo {author} {\bibfnamefont {E.}~\bibnamefont
  {R\"as\"anen}}, \bibinfo {author} {\bibfnamefont {M.~J.~T.}\ \bibnamefont
  {Oliveira}}, \bibinfo {author} {\bibfnamefont {D.~J.}\ \bibnamefont
  {Mowbray}}, \bibinfo {author} {\bibfnamefont {G.~M.}\ \bibnamefont
  {Nikolopoulos}}, \bibinfo {author} {\bibfnamefont {E.~T.}\ \bibnamefont
  {Karamatskos}}, \bibinfo {author} {\bibfnamefont {D.}~\bibnamefont
  {Markellos}}, \bibinfo {author} {\bibfnamefont {P.}~\bibnamefont
  {Lambropoulos}}, \bibinfo {author} {\bibfnamefont {S.}~\bibnamefont {Kurth}},
  \ and\ \bibinfo {author} {\bibfnamefont {A.}~\bibnamefont {Rubio}},\
  }\href@noop {} {\bibfield  {journal} {\bibinfo  {journal} {Phys. Rev. A}\
  }\textbf {\bibinfo {volume} {90}},\ \bibinfo {pages} {033412} (\bibinfo
  {year} {2014})}\BibitemShut {NoStop}%
\end{thebibliography}

%merlin.mbs apsrev4-1.bst 2010-07-25 4.21a (PWD, AO, DPC) hacked
%Control: key (0)
%Control: author (8) initials jnrlst
%Control: editor formatted (1) identically to author
%Control: production of article title (-1) disabled
%Control: page (0) single
%Control: year (1) truncated
%Control: production of eprint (0) enabled
%

\end{document}